\documentclass[11pt, a4paper]{article}
\usepackage{amsmath,amsthm, amssymb, latexsym}
\usepackage{mathtools}
\usepackage{amscd, amsfonts, cite, epsfig, hyperref, enumerate, graphics, subfig,epstopdf,float, color, bm}
\usepackage[left=2cm,top=2cm,right=2cm, bottom=1.95cm, nohead, a4paper]{geometry}

\usepackage{algorithmic}
\algsetup{linenosize=\small}
\usepackage{algorithm}

\begin{document}
\title{Maximum likelihood versus likelihood-free quantum system identification in the atom maser}

\author{  Catalin Catana\ \ \ Theodore Kypraios\ \ \ \ \ M\u{a}d\u{a}lin Gu\c{t}\u{a}\\
\normalsize{School of Mathematical Sciences, University of Nottingham}\\
\normalsize{University Park, Nottingham NG7 2RD, UK}}

\date{}

\maketitle

\vspace{8mm}

\begin{abstract}  

We consider the system identification problem of estimating a dynamical parameter of a Markovian quantum open system (the atom maser), by performing continuous time measurements in the system's output (outgoing atoms). Two estimation methods are investigated and compared. On the one hand, the maximum likelihood estimator (MLE) takes into account the full measurement data and is asymptotically optimal in terms of its mean square error. On the other hand, the `likelihood-free' method of approximate Bayesian computation (ABC) produces an approximation of the posterior distribution for a given set of summary statistics, by sampling trajectories at different parameter values and comparing them with the measurement data via chosen statistics.

\vspace{2mm}


Our analysis is performed on the atom maser model, which exhibits interesting features such as bistability and dynamical phase transitions, and has connections with the classical theory of hidden Markov processes. Building on previous results which showed that atom counts are poor statistics for certain values of the Rabi angle, we apply MLE to the full measurement data and estimate its Fisher information. We then select several correlation statistics such as waiting times, distribution of successive identical detections, and use them as input of the ABC algorithm. The resulting posterior distribution follows closely the data likelihood, showing that the selected statistics contain `most' statistical information about the Rabi angle.

\end{abstract}

\section{Introduction}

System identification is a field lying at the overlap between control theory and statistical inference \cite{Ljung}. The typical system identification problem is that of estimating certain parameters of a dynamical system by designing input signals, while monitoring and analysing the corresponding output signals. Similar problems arise in quantum engineering, where the task may be to estimate a quantum channel \cite{Fujiwara}, the Hamiltonian of a quantum system \cite{Burgarth,Cole}, or the Lindblad generator of an open dynamical system in the Markov approximation \cite{Howard}.  
By analogy, we expect that quantum system identification will play an important role in quantum control  theory \cite{Mabuchi&Khaneja} and the development of new quantum technologies \cite{Dowling&Milburn}.

This paper deals with the following scenario, which is distinct from the channel estimation set-up. An open quantum system interacts weakly with the environment which is monitored by means of continuous time measurements; the measurement output is used to estimate a dynamical parameter  \cite{Mabuchi}, for instance the system-environment interaction strength. In the Markov approximation, the joint dynamics is described in the quantum input-output formalism \cite{Gardiner&Zoller,Wiseman&Milburn}, by means of quantum stochastic differential equations \cite{Parthasarathy}. The output measurement process carries information about the dynamics and can be analysed using statistical inference tools. In particular, using the quantum trajectories formalism \cite{Gardiner&Zoller,Belavkin}  one can compute  the likelihood of a measurement trajectory, and estimate the the unknown parameter by applying e.g. maximum likelihood, or  Bayesian inference methods.

The discrete-time version of this set-up was analysed in \cite{Guta} which shows that  both the output quantum state, and the average statistics of repeated measurements are asymptotically normal (Gaussian), and provides explicit expressions for the corresponding quantum and classical Fisher informations (cf. \cite{Guta&Kiukas} for a more general treatment). In continuous-time, a similar analysis was undertaken in  \cite{Catana} for the particular model of the atom maser: a cavity interacting with incoming identically prepared two-levels atoms which are subsequently measured to produce a continuous-time counting process, as illustrate in Figure \ref{fig.atommaser}. One of the intriguing findings was that for certain values of the unknown parameter (the Rabi angle $\phi$),  the total atom counts are poor statistics (zero classical Fisher information), while the \emph{quantum} Fisher information of the output attains its maximum.

Here we extend this analysis in two directions. Firstly, we apply the maximum likelihood estimation method (MLE) to the full atom detection record, rather than the total counts statistics as in \cite{Catana}. 

This is done by computing the likelihood of a detection record using the quantum trajectories formalism. 
The estimated classical Fisher information is illustrated in Figure \ref{fig:mle} in comparison with that of the total counts statistics. The significant discrepancy between the two informations in the region around $\phi\approx 0.4$ motivated a further investigation aimed at finding more informative statistics, and other estimation methods which may be less computationally demanding than MLE.

With this in mind, in the second part of the paper we introduce and analyse a set of 7 statistics such as waiting times between successive detections of a given type (ground or excited state), number of successive detections of the same type, and density of detections of a given type before a detection of the other type. A strong dependence on $\phi$ of the statistic's distribution indicates that it is informative about the parameter. However, likelihood-based estimation for such statistics is typically as involved as performing MLE on the full measurement data. Therefore we employ a `likelihood-free' method known as approximate Bayesian computation (ABC) which works as follows (cf. section \ref{sec: ABC} for details). A large number of measurement trajectories are \emph{simulated} for each parameter value chosen from a given interval with uniform probability, and each sample is accepted if the value of the corresponding statistic is `sufficiently close' to that of the same statistic for the `measurement data'. 
The histogram of accepted samples is an approximation of the posterior distribution of $\phi$ given the summary statistic(s) used in the ABC procedure, and the accuracy can be controlled by a rejection threshold parameter. 

When applied to \emph{individual} statistics the method produces posterior distribution with significantly larger variance that the MLE. However, when  \emph{all} statistics are taken into account simultaneously, the ABC performance is very close to that of the MLE of the full measurement data, as illustrate in Figure \ref{fig:alltogether}. This shows that `most' information about $\phi$ is captured by a small number of appropriately chosen statistics, which is one of the main insights of the paper. We therefore anticipate that likelihood-free methods such as ABC can be a valuable alternative to standard MLE in models where the likelihood function is unknown or difficult to compute, while sampling the measurement process at given parameters is relatively easier.

In section \ref{atom.maser} we introduce the atom maser model and the quantum trajectories formalism, and present an algorithm for simulating the different stochastic processes. Section \ref{sec.statistical.inference} gives a brief overview of the statistical concepts employed in the paper: Fisher information for independent data and Markov processes, asymptotic normality, the maximum likelihood estimator, the ABC method for producing an approximate posterior distribution. Section \ref{sec.results} contains the main results of the paper. We discuss several measurement scenarios and their associated classical Fisher informations, in particular the Fisher information of the total number of atom counts statistics is compared to that of the full atoms measurement data, cf. Figure \ref{fig:mle}. We then consider several atom measurements statistics, such as waiting times,  number of successive detections of the same type, total number of detections of the same type, and apply the ABC procedure for each of them 
separately, and jointly for several statistics. 

\section{The Atom Maser}
\label{atom.maser}

\begin{figure}[H]
    \begin{center}  
       \includegraphics[width=0.7\textwidth]{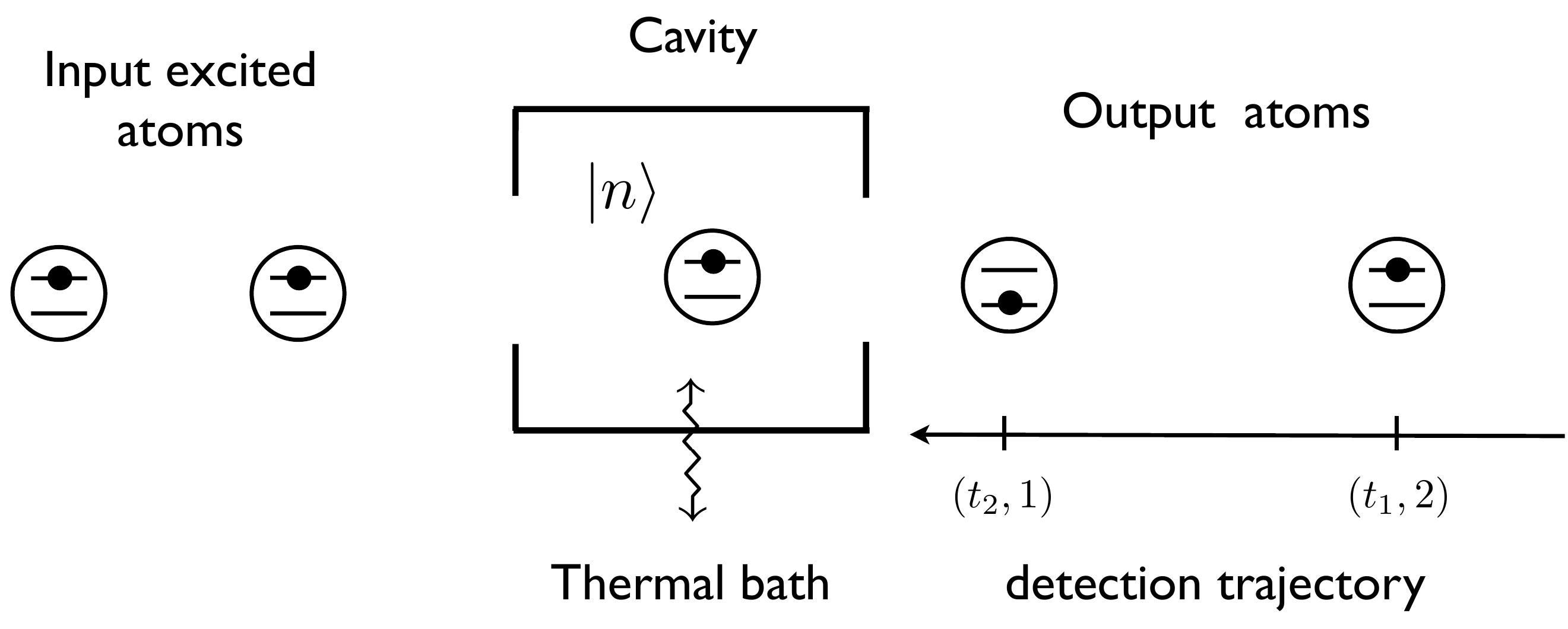}
   \caption{Atom maser: two level atoms in the excited state are pumped through a dissipative  cavity with Poisson distribution of rate $N_{ex}$. The outgoing atoms are measured in standard basis producing a detection record $\{(t_1, i_1), \dots ,(t_k, i_k). \}$}
\label{fig.atommaser}
 \end{center}
\end{figure}

We consider the standard atom maser model \cite{Meschede,Englert2} which consists of a beam of two-level atoms passing through and interacting resonantly  with the electromagnetic field of a dissipative cavity, cf. Figure \ref{fig.atommaser}. The incoming atoms are prepared in the excited state, have equal velocities and their arrival times are Poisson distributed with a given rate $N_{ex}$. The atom-field interaction is of Jaynes-Cummings type \cite{Jaynes}, and  the cavity is in contact with a low temperature thermal bath 
such that the field can both emit or absorb photons with given rates. Following \cite{Englert2} we assume that at any moment of time at most one atom is present in the cavity. This leads to a coarse grained continuous time master evolution of the cavity as described below. The atoms emerging from the cavity are detected and measured in the standard basis. This continuous-time measurement produces a record of successive detection times, each time carrying the label  of the measurement outcome ($1$ for ground and $2$ for excited state).

The goal is to infer the  unknown value of the atom-field interaction strength, or more precisely the accumulated Rabi angle, from the stochastic measurement record on a long time interval. The key differences with the channel tomography set-up are that we do not measure the system (cavity) directly, and we do not perform repeated independent measurements, but use a single-shot, time-correlated measurement record to infer the unknown parameter. For sufficiently long times, the cavity reaches its 
unique steady state, and the measurement process is stationary. In this regime one can compute the Fisher information for certain \emph{statistics} of the measurement data and prove their asymptotic normality \cite{Catana}. Here we complete this analysis by investigating and comparing the maximum likelihood estimator based on the \emph{full} measurement record, and the alternative \emph{likelihood-free} method of approximate Bayesian computation, based on a set of measurement statistics. 

Although it may be practically unrealistic, for comparison purposes it is useful to consider thought experiments where in addition to the atoms measurement, the photons emitted and absorbed by the cavity, are also detected. Other scenarios that will not be considered here but may be investigated in a similar fashion, include the detection of outgoing atoms in different basis (phase-sensitive measurements) \cite{Englert}, imperfect measurements, and atom masers with different pumping statistics \cite{Guerra,Herzog,Bergou}. 

\subsection{Master dynamics}

Assuming that the time scale of the atom-cavity interaction is much smaller than both the decay time of the field and the typical inter-arrival time of the atoms, the cavity dynamics is governed by the master equation  \cite{Englert2,Englert3}
\begin{equation}\label{eq:lindblad}
\frac{d\rho}{dt} = \mathcal{L} (\rho)= \sum_{i=1}^{4}\left( L_{i} \rho L_{i}^{\dagger} - \frac{1}{2}\lbrace L_{i}^{\dagger} L_{i}, \rho \rbrace \right),
\end{equation}
where  $L_{i}$ are the four operators 
\begin{equation*}
 L_{1} = \sqrt{N_{ex}} a^{\dagger} \frac{\sin(\phi \sqrt{a a^{\dagger}})}{\sqrt{a a^{\dagger}}}, \quad L_{2} = \sqrt{N_{ex}} \cos(\phi \sqrt{a a^{\dagger}}), \quad L_{3} = \sqrt{\nu+1}a, \quad L_{4} = \sqrt{\nu} a^{\dagger},
\end{equation*}
corresponding to the possible quantum jumps. The first two describe jumps induced by the detection of an atom in the ground and respectively excited state, while the last two are due to photon emission and absorption, with $a^{\dagger}$and $a$ the creation and annihilation operators of the cavity field, and $\nu$ the photon absorption rate. The unique stationary state of the master dynamics satisfies $\mathcal{L}(\rho^{ss})=0$, and the density matrix 
$\rho^{ss}$ is diagonal in the Fock basis, with coefficients  given by
\begin{equation*}
\rho^{ss}(n)=\rho^{ss}(0)\prod_{i=1}^{n}{\left[\frac{\nu}{\nu+1}+\frac{N_{ex}}{\nu+1}\frac{\sin^{2}(\phi \sqrt{i})} {i}\right]}.
\end{equation*}

\begin{figure}[H]
    \begin{center}  
       \includegraphics[width=0.5\textwidth]{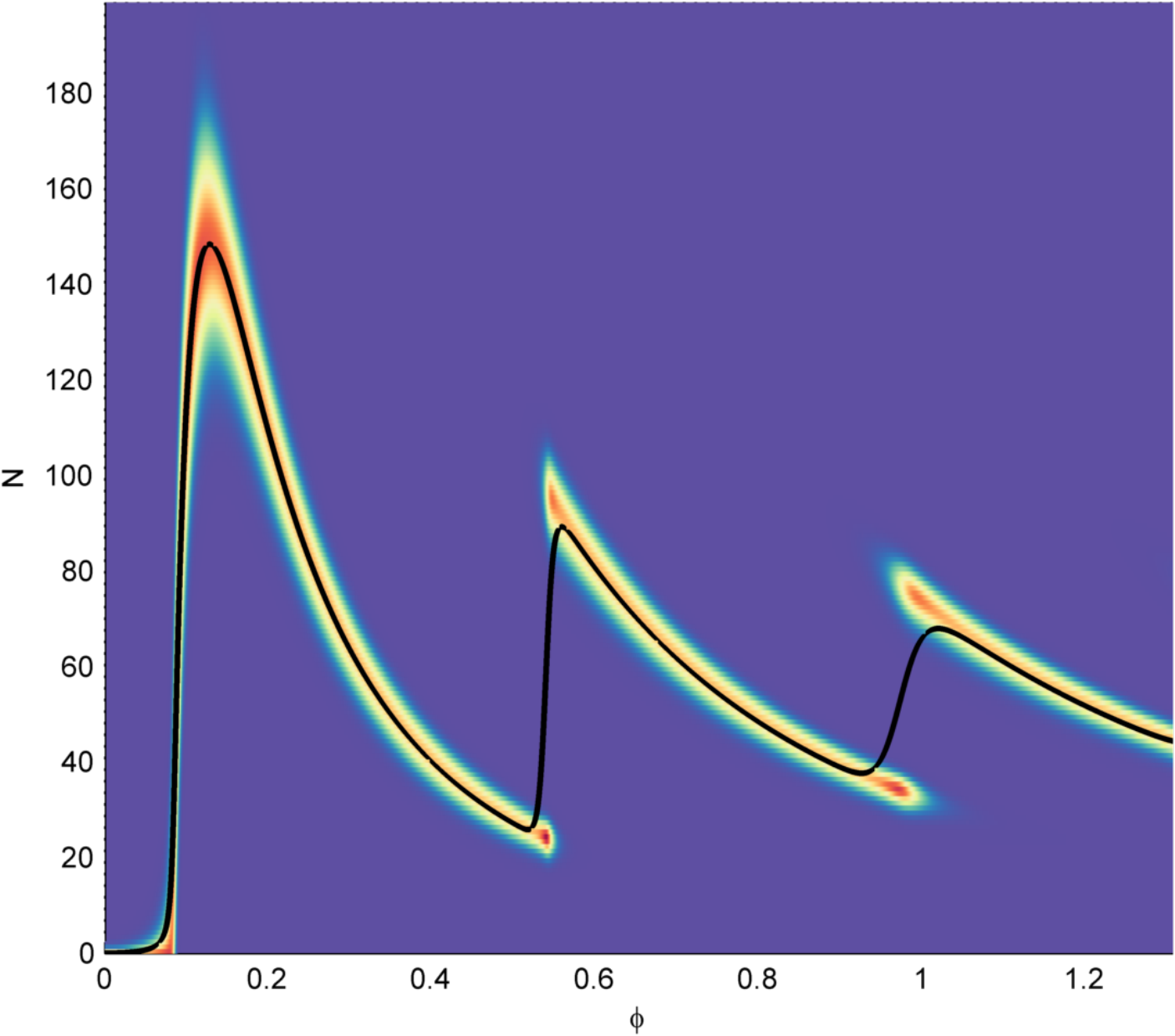}
   \caption{The stationary distribution of photon numbers (colored patches) is represented as a function of the Rabi angle $ \phi$ at $\nu=0.15$ and $N_{ex}=150$. The black curve represents the cavity mean photon number.}
\label{fig.stationarystate}
 \end{center}
\end{figure}

Figure \ref{fig.stationarystate} shows the stationary photon number distribution and the mean photon number, as a function of the Rabi angle $\phi$. The distribution has several features which are relevant for our investigation. Firstly, it is multi-stable at certain values of $\phi$, e.g. it exhibits bistability for $0.51<\phi<0.61$ and the mean photon number changes abruptly around $\phi=0.55$. As we will see below, this is reflected in the fact that the quantum trajectories alternate between long periods of high and low photon numbers corresponding to active and passive periods of ground-state atoms detections. 
Secondly, the mean photon number peaks at 
$\phi\approx 0.18$, so that parameters which are close to this point and are situated symmetrically with respect to it have similar mean photon numbers and are harder to distinguish statistically. At very low temperatures the  stationary state  exhibits other interesting features such as trapping \cite{Meystre}.

\subsection{Quantum trajectories}
\label{sec:detection}

From the master dynamics level, we pass to that of quantum trajectories which describes the stochastic evolution of the cavity conditional on the atom (or photon) detection events. Generally, quantum trajectories are solutions of stochastic Schr\"{o}dinger equations \cite{Bouten}, or filtering equations \cite{Belavkin} driven by the \emph{stochastic} measurement process; the \emph{deterministic} dynamics on the master level can then be recovered by averaging over measurement trajectories. In this work, the quantum trajectories are mainly used as a tool for investigating the statistical properties of the measurement process, but we envisage that they would play a more significant role in a scenario where the measurement strategy is optimised and includes feedback control.

 During the interaction with the cavity the atoms are entangled with the cavity field, and are subsequently detected  in either the ground or the excited state. We assume an ideal measurement process, i.e. all atoms are detected, and each atom undergoes a von Neumann projective measurement, and we refer to \cite{Sterpi} for a more general set-up including undetected atoms and measurement errors. 
We will start by describing a `full environment monitoring' scenario where we assume that besides the atoms,  the emitted and absorbed photons are also detected. After that we will focus on the more realistic situation where only the atoms are detected, which can be seen as `tracing over' the photon detection events.

\begin{figure}[H]
  \hfill
  \begin{minipage}[t]{.45\textwidth}
    \begin{center}  
       \includegraphics[width=0.99\textwidth]{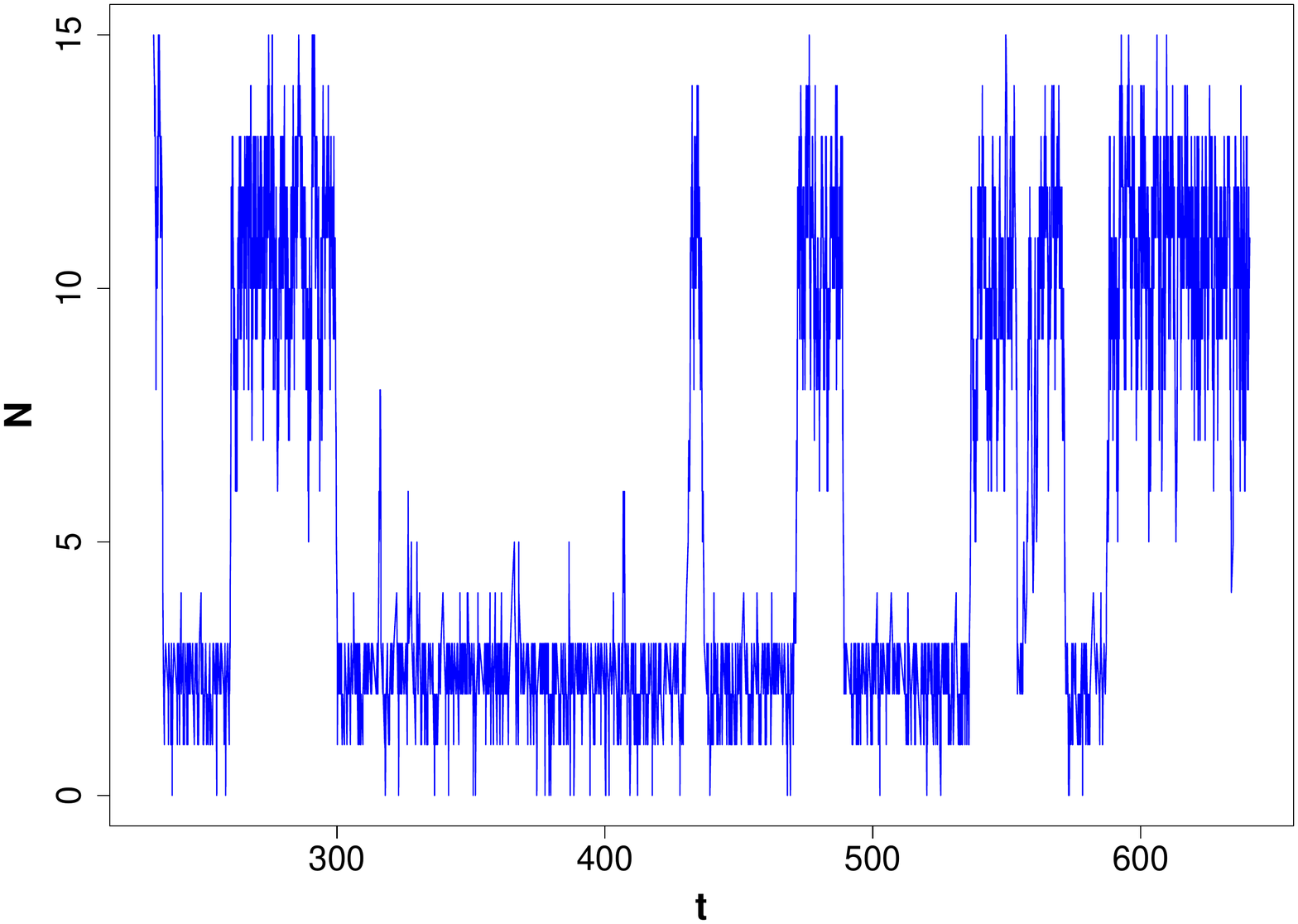}
    \end{center}
  \end{minipage}
  \hfill
  \begin{minipage}[t]{.45\textwidth}
    \begin{center}  
  \includegraphics[width=0.99\textwidth]{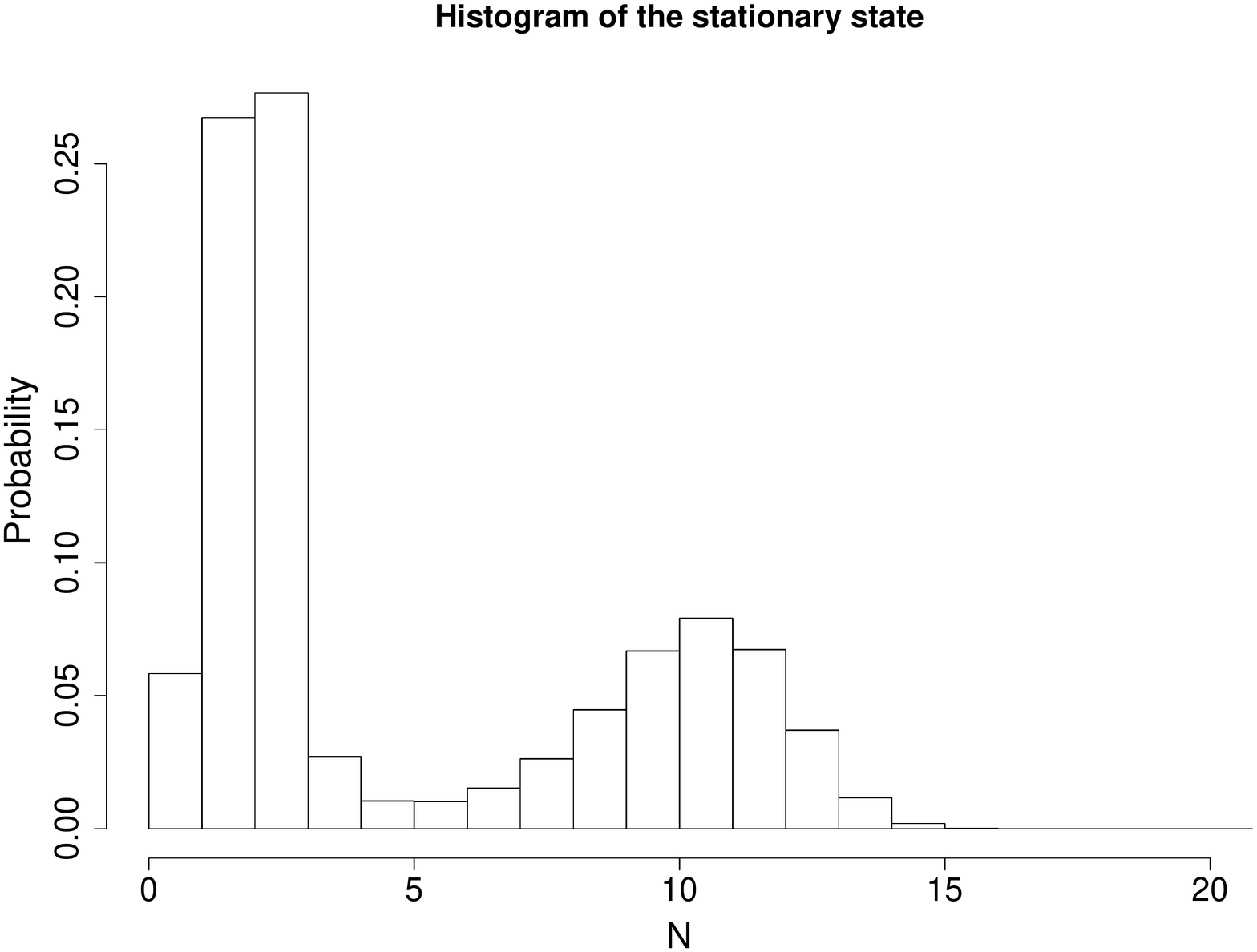}
    \end{center}
  \end{minipage}
  \hfill
\caption{In the full environment monitoring scenario, the cavity state evolves as a birth and death process on the Fock ladder. Left panel: when the stationary state is bistable ($\phi=1.6$, $N_{ex}=16$) the state spends long periods in  `phases' with distinct mean photon numbers, with quick `jumps' between the phases. Right panel: histogram of the total number of ground-state atoms (per unit of time) is bistable, exhibiting active and passive dynamical phases.}
\label{fig:bistability}
\end{figure}

Whenever an atom is detected in the ground state, the cavity performs a 'quantum jump' from the current state 
$\rho(t)$ to the new state  
\begin{equation}\label{eq:state.reduction}
\rho^\prime(t):= \frac{\mathcal{J}_1\rho(t)}{{\rm Tr}\left\{\mathcal{J}_1 \rho(t)\right\}}, \qquad
\mathcal{J}_1 \rho:=L_{1}\rho L_{1}^{\dagger}.
\end{equation}
Similarly, we define the jump superoperators $\mathcal{J}_2,\mathcal{J}_3,\mathcal{J}_4$ corresponding to the detection of a ground-state atom, the emission and absorption of a photon. Between two successive detections the cavity state evolves continuously as   
$$
\rho(t+\tau) := \frac{e^{\tau  \tilde{\mathcal{L}}} \rho(t)}{{\rm Tr} \{e^{\tau  \tilde{\mathcal{L}}} \rho(t)\}} , \qquad \tilde{\mathcal{L}} \rho:= - \frac{1}{2}\sum_{i=1}^4\left\{ L_i^{\dagger}L_i, \rho \right\},
$$
which is equivalent to an evolution with the `effective Hamiltonian' $H_0=- i/2 \sum_{i=1}^4 L_i^{\dagger}L_i$, on the level of pure states. A general feature of continuous-time measurements where the environment is fully monitored, is that the system dynamics can be run solely on pure states, thus reducing the overhead in computer simulations. In the case of the atom maser, the complexity can be further reduced by noticing that if the cavity is initially in a Fock state, it will always jump up or down the Fock ladder in a classical fashion. The corresponding stochastic dynamics is that of a birth and death process with rates \cite{Catana}
\begin{eqnarray}
&&
q_{k,k+1}:= N_{ex} \sin(\phi \sqrt{k+1})^{2} + \nu (k+1) ,\quad k\geq 0\nonumber \\
&&
q_{k,k-1} := (\nu+1)k, \quad k\geq 1.\label{eq.birth-death}
\end{eqnarray}
Figure \ref{fig:bistability} shows a particular realisation of this dynamics where the pumping rate has been chosen such that the stationary state is bistable. The quantum trajectory (left panel) exhibits long periods of evolution inside either the high or the low photon 'phases', with quick 'jumps' between them. This behaviour is mirrored by the total number of ground-state atoms statistic which alternates between active and passive 'dynamical phases' \cite{Igor,Merlijn,Benson}. 

We consider now the second, more realistic scenario where only the atoms are detected in one of the two states. Since in this case the dynamics is obtained by averaging over all unobserved photon events, the cavity state will be generally mixed and needs to be described by a density matrix. The jump superoperators $\mathcal{J}_1$ and $\mathcal{J}_2$ are the same as above, but the the evolution between jumps has generator $\mathcal{L}_0$ which replaces $\tilde{\mathcal{L}}$, and takes into account all photon events 
$$
\mathcal{L}_0 \rho:=L_{3}\rho L_{3}^{\dagger}+L_{4}\rho L_{4}^{\dagger}-\frac{1}{2}\left\{ L_3^{\dagger}L_3+ L_4^{\dagger}L_4, \rho \right\}-N_{ex} \rho.
$$
Let $t_{1}<t_{2} < \dots $ be a sequence of detection events occurring after the initial time $t_0$, and let $i_{k}\in \{1,2\}$ be the jump type for the atom detected at time $t_{k}$.  We denote by ${\bf d}_{t]}=\{ (t_1, i_1), \dots, (t_n, i_n)\}$ the detection record up to time $t$, with $t_n\leq t<t_{n+1}$. If the initial state of the field is $\rho_{0}$, then the corresponding (un-normalised) quantum  trajectory at time $t$ is given by  
\begin{equation} \label{eq:condstate}
\rho(t~;~{\bf d}_{t]})= 
e^{\mathcal{L}_{0}(t-t_{n})} \mathcal{J}_{i_{n}} e^{\mathcal{L}_{0}(t_{n}-t_{n-1})}\mathcal{J}_{i_{n-1}}\dots 
\mathcal{J}_{i_{1}} e^{\mathcal{L}_{0}t_{1}} \rho_{0}.
\end{equation}
The corresponding probability density is equal to  
$
p({\bf d}_{t]}) = {\rm Tr}(\rho(t~;~ {\bf d}_{t]})),
$
and satisfies the normalisation condition
$$
\sum_{n=0}^\infty ~\sum_{i_1,\dots ,  i_n \in \{1, 2\}} ~
\int_{t_0}^t\dots \int_{t_0}^t p({\bf d}_{t]}) dt_1\dots dt_n =1.
$$

\subsection{Simulations}
\label{sec.simulations}

In this section we briefly describe how to simulate the process which describes the stochastic evolution of the cavity (see Section \ref{sec:detection}) in the time interval $[0, \tau]$. We are primarily interested in the long time behaviour when the dynamics looses memory of the initial cavity state and therefore, we can assume that its initial state is the stationary state $\rho^{ss}$. Moreover, since the jump operators preserve the set of diagonal states, the probability $p({\bf d}_{t]})$ can be computed by restricting the attention to such states, rather than full density matrices. This reduces the problem to that of simulating a birth and death process (the cavity state) together with two processes describing the atoms counts. 

Denote by $ N(t), \Lambda_1(t), \Lambda_2(t)$ the $\mathbb{N}$-valued stochastic processes which describe the cavity state at time $t$, the total number of ground-state and excited-state atoms detected up to time $t$ respectively. The main steps of the algorithm are as follows.
\vspace{0.1in}
\begin{algorithm}[H]

 {\bf 1.} Set $t=0$ and initialise $N(t)$ by drawing randomly from $\rho^{ss} (\cdot)$  i.e. $N(0) \sim \rho^{ss} (\cdot)$;

 {\bf 2.} Simulate a birth and death process $\{N(t), \, t \in [0, \tau]\}$ with rates given in Equation \eqref{eq.birth-death};

 {\bf 3.} Simulate a process which determines if the absorption of a photon was due to an atom or the bath;

 {\bf 4.} Simulate a Poisson process between two consecutive jumps of $\{ N(t) \}$ whose event times determine the detection times of the excited atoms.
\caption{\newline Simulation of counting processes}
\label{alg:simulation}
\end{algorithm}
We now explain in detail how each of the steps is performed:

\begin{itemize}
\item Step 2 describes the evolution of the cavity state in the full environment monitoring scenario. Let $s_1< s_2<\dots$ denote the (random) times at which the cavity state makes a jump and $N_k$ be the state of the cavity after the $k$'th jump. Define $J_k:=N_k-N_{k-1}\in \{-1,+1\} $ to be the increment indicating which type of jump (up or down) has occurred. 

\item Step 3 is carried out as follows: if $J_k=+1$ then the cavity has absorbed a photon which can either come from an excited input atom or from the bath. To decide which of two type of events has occurred we draw  $C_k \sim \mbox{Bernoulli}(p)$ with probability of a success (i.e. jump was due to an atom) equal to $p=\nu (N_k +1 )/\left(N_{ex} \sin\left(\phi \sqrt{N_k +1}\right)^{2} + \nu (N_k +1)\right)$. In addition, we increment $\Lambda_1(t)$ by one. If $J_k=-1$ the cavity has emitted a photon in the bath, and no atom detection needs to be recorded. 

\item In Step 4 the arrival times of the excited--state atoms in each time interval $[s_k, s_{k+1}]$ are simulated according to the event times of a Poisson process with state--dependent intensity $N_{ex} \cos\left(\phi \sqrt{N_k+1}\right)^{2}$. Note that the Poisson processes for different intervals are asssumed to be independent and that the number of excited-atoms $\Lambda_2(t)$ is equal to the to total number of arrival times up to $t$. 
\end{itemize}

\begin{figure}[H]
   \begin{center}  
 \includegraphics[width=0.99\textwidth]{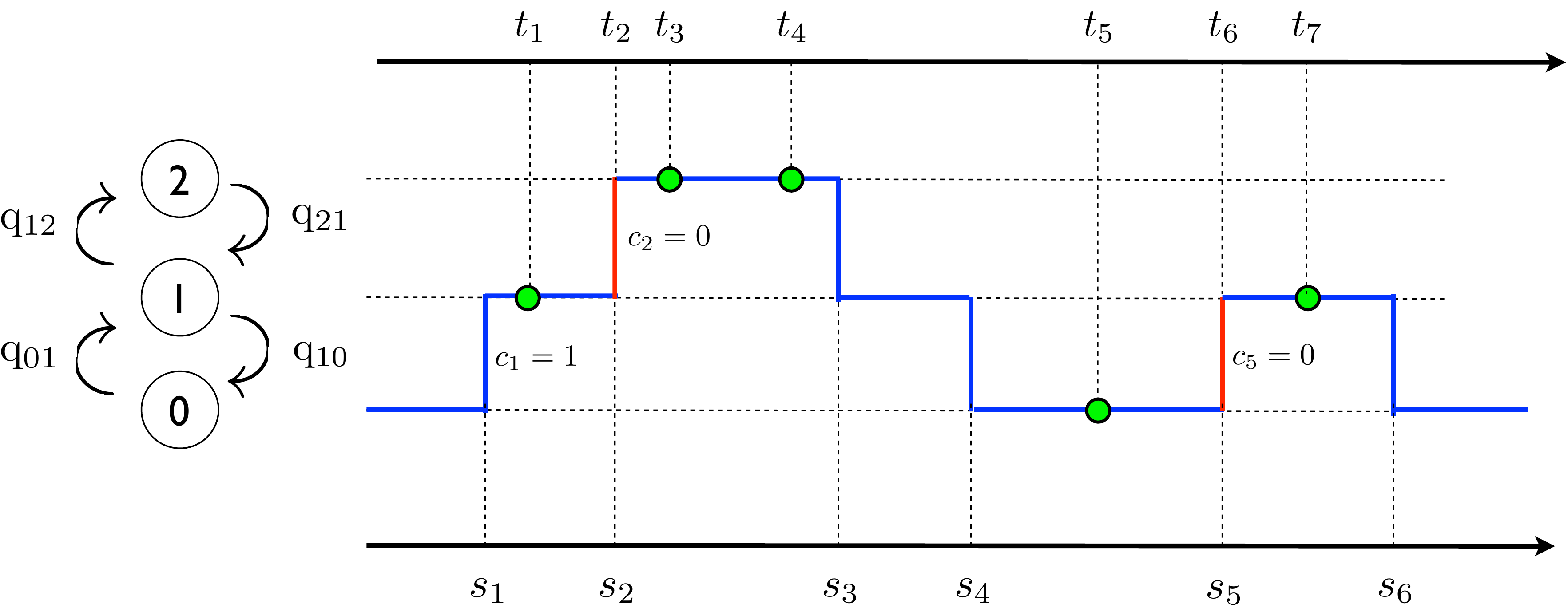}
   \end{center}

\caption{Stochastic processes derived from the atom maser dynamics. The cavity is described by a birth and death process (blue curve) with jump rates $q_{k,k+1}$ and $q_{k,k-1}$. When the cavity absorbs a photon, a coin is tossed to decide whether this was due to an atom (red vertical lines) or the bath (blue vertical lines). On the time intervals between two jumps, we generate a Poisson process (green dots) with state dependent intensity, representing the detection times of excited state atoms.} 
\label{fig:simulations}
\end{figure}

Finally, the arrival times of both ground and excited state atoms are collected in the sequence $t_1<t_2< \dots$, and the type of atom arriving at $t_k$ is encoded in the label $i_k\in \{1,2\}$. Figure \ref{fig:simulations} illustrates the above algorithm.

\section{Classical and quantum statistical inference}
\label{sec.statistical.inference}
 
In this section we review the basic notions and methods of statistical inference employed in the present paper. We begin with an introduction to statistical inference for classical and quantum systems and then review a particular likelihood-free inference method used in Bayesian statistics.

\subsection{Basic notions of frequentist inference}
\label{sec.frequentist}

Our starting point is the following basic statistical problem: given $n$ independent identically distributed (i.i.d.) 
samples $X_1,\dots ,X_n\in \mathcal{X}$ drawn from a probability distribution $\mathbb{P}_\theta$ which depends on  an unknown parameter 
$\theta\in \Theta\subset\mathbb{R}^p$, we would like to construct an estimator $\hat{\theta}_n:=\hat{\theta}_n(X_1,\dots ,X_n) $ such that the `size' of the estimation error $\hat{\theta}_n- \theta$ is as small as possible. The Cram\'{e}r-Rao (CR) inequality states that the covariance matrix of any \emph{unbiased} estimator satisfies the lower bound
$$
n\mathbb{E}\left[ (\hat{\theta}_n-\theta)^T(\hat{\theta}_n-\theta)\right] \geq I(\theta)^{-1}
$$
where $I(\theta)$ is the Fisher information matrix (for one sample) defined as
\begin{equation}\label{eq.Fisher}
I(\theta)_{ij}=\int_{p_\theta(x)>0} p_\theta(x) \frac{\partial \log p_\theta(x)}{\partial \theta_i}\frac{\partial \log p_\theta(x)}{\partial \theta_j} dx,
\end{equation}
with $p_\theta$ the density of $\mathbb{P}_\theta$ with respect to some measure $dx$ on $\mathcal{X}$. The importance of the CR bound lies in the fact that it is \emph{asymptotically achievable} in the sense that there exist good (or efficient) estimators which satisfy the \emph{asymptotic normality} property 
\begin{equation}\label{eq.asymptotic.normality.ml}
\lim_{n \to \infty}\sqrt{n}(\hat{\theta}_n-\theta)\overset{\mathcal{L}}{\longrightarrow} N(0, I(\theta)^{-1}),
\end{equation}
where the limit  convergence in law to the centred normal distribution $N(0,I(\theta)^{-1})$.
In particular, under certain regularity conditions the maximum likelihood estimator 
$$
\hat{\theta}_n = \arg\max_{\theta^\prime\in \Theta}\, \sum_i \log p_{\theta^\prime}(X_i)
$$ 
is an efficient estimator \cite{Young&Smith}, which explains it popularity as a statistical estimation tool. 

\vspace{4mm}

Asymptotic normality is a general phenomenon which also holds for models with dependent data, such as a (hidden) Markov process. Let $(X_n)_{n\in \mathbb{N}}$ be an ergodic Markov chain  with discrete states space $I:=\{ 1,\dots , m\}$, transition matrix $T=(T_{i,j})_{i,j\in I}$, and stationary distribution $\pi:= (\pi_1, \dots, \pi_m)$. As above, suppose that $T$ depends on an unknown parameter $\theta\in \Theta\subset\mathbb{R}^p$ and we would like to estimate $\theta$ from the  trajectory $X_1, \dots, X_n$. In the long time limit the process reaches stationarity, and the associated maximum likelihood estimator $\hat{\theta}_n$ is again asymptotically normal, i.e. the limit \eqref{eq.asymptotic.normality.ml} holds with the corresponding Fisher information (per unit of time) given by 
\begin{equation}\label{eq.fisher.markov}
I\left(\theta\right):=\sum_{i\neq j}\pi_{i}^{\theta} T_{ij}^{\theta} \left(\ell_{ij}^{\theta}\right)^{2}, \qquad 
\ell_{ij}^{\theta}:=\frac{d}{d\theta}\log T_{ij}^{\theta}
\end{equation}
where $\pi^{\theta}$ is the stationary distribution at $\theta$.

Hidden Markov processes form an important class of statistical models extending the notion of Markov process. Here we assume that the above Markov chain $(X_n)_{n\in\mathbb{N}}$ is not directly accessible, and the observations consist of another process $(Y_n)_{n\in \mathbb{N}}$ with values in $J= \{ 1,\dots, k\}$ such that $Y_1, \dots , Y_n$ are independent conditionally on $X_1,\dots, X_n$ and $Y_i$ depends only on $X_i$ with conditional distribution given by
$$
Q_{i,j} = \mathbb{P}(Y_i = y | X_i = x ). 
$$ 
The identifiability of the matrices $T$ and $Q$ is discussed in \cite{Petrie} and the asymptotic normality of the MLE is discussed in \cite{Baum} and in \cite{Ryden2,Douc,Leroux} for more general hidden Markov models. In this case the limiting Fisher information does not have a simple expression but it can be estimated from the data by the \emph{observed} Fisher information given by 
$$
\hat{I}_n = -
\frac{1}{n}
\left.\frac{\partial^{2}\ell_\theta  \left( Y_1, \dots , Y_n\right) }{ \partial \theta ^{2}} \right|_{\theta=\hat{\theta}_n}, \qquad
\ell_\theta  \left( Y_1, \dots , Y_n\right) =\log p_\theta (Y_1, \dots , Y_n)
$$
where $\hat\theta_n$ is the maximum likelihood estimator and $p_\theta (Y_1, \dots , Y_n)$ is the likelihood function for data 
$Y_1, \dots , Y_n$. This formula will be used later for estimating the Fisher information for the atom detection process in the atom maser.

\subsection{Quantum statistics}

In this section we give a very brief review of the key notions of quantum statical inference that are relevant for the paper. 
For more background reading we refer to \cite{Holevo,Gill,Gill&Guta}.

\subsubsection{Quantum state estimation}
The standard quantum statistical problem is that of state estimation or quantum tomography. 
Given a number of independent copies of a state $\rho^\theta$ depending on an unknown parameter 
$\theta\in \Theta$, we would like to estimate the parameter by performing (separate or joint)  measurements  on the quantum systems and use the results to construct an estimator of $\theta$.

If all measurements are identical and are characterised by the positive operator valued measure (POVM) $M(dx)$ over a space $\mathcal{X}$, then the measurement outcomes are i.i.d. with probability distribution
$$
\mathbb{P}^M_\theta(dx) = {\rm Tr} ({\rho}_\theta M(dx)).
$$
The model $\{ \mathbb{P}^M_\theta : \theta\in \Theta\}$  has an associated \emph{classical Fisher information} $I(\theta; M)$ given by  \eqref{eq.Fisher}. 

If $\theta$ is one-dimensional, the optimal measurement can found by maximising $I(\theta;  M)$ over all 
measurements $M$. A solution of this optimisation is the measurement of the observable $\mathcal{L}_\theta$ called \emph{symmetric logarithmic derivative} defined by the equation
$$
\frac{d\rho_\theta}{d\theta}= \frac{1}{2}\left (\mathcal{L}_\theta \rho_\theta + \rho_\theta \mathcal{L}_\theta  \right ).
$$
The associated information is called the \emph{quantum Fisher information}, it depends only on the intrinsic properties of the quantum statistical model $\{\rho_\theta :\theta\in \Theta\}$ and is equal to \cite{Holevo,Caves}
$$
F(\rho_\theta)={\rm Tr}\left \{  \rho_\theta \mathcal{L}_\theta^2\right\}. 
$$ 
Since $\mathcal{L}_\theta$ depends on the unknown parameter $\theta$, in practice one uses an adaptive two step procedure, where a small proportion of the systems are used to produce a rough estimate $\theta_0$ of $\theta$, and the observable $\mathcal{L}_{\theta_0}$ is subsequently measured on the rest of the samples. Asymptotically, this procedure produces 
(maximum likelihood) estimators which are asymptotically normal, and achieve the quantum Fisher information in the sense of \eqref{eq.asymptotic.normality.ml}.

If $\theta:= (\theta_1, \dots, \theta_k)$ is multidimensional, then the quantum Fisher information can be defined in a similar way, using the SLD's
$\mathcal{L}_{\theta,i}$ corresponding to the coordinate $\theta_i$. However, since these observables may not commute with each other, the quantum Fisher information may not be achievable by any measurement. One can find the asymptotically optimal solution for a given loss function (figure of merit) which is locally quadratic in $\theta$, but in general the optimal measurements for different loss functions are incompatible with each other \cite{Guta}. A theory extending the concept of asymptotic normality to the quantum set-up has been developed in \cite{Guta&Kahn,Guta&Jencova}, where it is shown how the $n$ samples model converges to a quantum Gaussian model with unknown mean and fixed covariance matrix. In this way the original estimation problem can be transformed into the simpler problem of estimating 
the mean of a Gaussian state.

\subsubsection{System identification for quantum Markov processes}

In this paper we are concerned with the estimation of a one-dimensional parameter. However the quantum statistical model does not consist of identically prepared, independent systems but is the (continuous time) output of a quantum Markov process. As in the case of classical Markov processes, the `data' is not iid but carries correlations due to the memory carried by the system.


The discrete time (ergodic) quantum Markov chain set-up was  investigated in \cite{Guta}, and a more general treatment will be presented in\cite{Guta&Kiukas}. For concreteness we can think of the atom-maser model illustrated in Figure \ref{fig.atommaser}, with atoms passing at equal time intervals  and the interaction between atoms and system (cavity) given by a joint unitary transformation $ U^\theta$ depending on the unknown parameter $\theta$. Two extreme scenarios can be considered. On the one hand, the experimenter can perform identical repeated measurements on the output, and use the \emph{average} of the results to estimate the unknown parameter. Such additive statistics are asymptotically normal and the corresponding Fisher information has an explicit expression. The other extreme is to consider the output as a quantum system whose state depends on the unknown parameter, and therefore has a quantum Fisher information. It turns out that the state itself is asymptotically Gaussian in the statistical 
sense defined in \cite{Guta&Kahn,Guta&Jencova} and the corresponding 
(asymptotic) quantum Fisher information has an explicit expression. 

These results can be extended to continuous time \cite{Catana&Bouten&Guta}, and the particular case of the atom maser was investigated in \cite{Catana}. However the analysis in the latter work was based solely on the total counts statistics and did not take into account the full atom measurement data, cf. section \ref{sec.likelihood.atom.maser} for more details. In the present work we fill this gap, and complement and compare it with a likelihood-free method which uses a small number of statistics of the data, such as waiting time distribution and average number of successive detections of the same type. 
The details of this method are described in the next section.

\subsection{Approximate Bayesian Computation}\label{sec: ABC}

Similarly to the frequentist setting, Bayesian inference requires a sampling space $\mathcal{X}$ and a likelihood 
$p(x|\theta)$, which can be seen as the conditional distribution of the data $X$ given the parameter $\theta$. Additionally, the Bayesian approach places a prior distribution with density $\pi(\theta)$ on the model parameters, which encodes our prior beliefs prior to seeing the data. The likelihood and the prior are then combined using Bayes theorem to derive the posterior distribution which essentially is the conditional distribution of the (unknown) parameter $\theta$, given the data $X\in \mathcal{X}$ 

\begin{equation}
 p(\theta| X) = \frac{p(X|\theta) \pi(\theta)}{\int_{\theta} p(X|\theta)\pi(\theta) \mbox{d}\theta} = \frac{p(X|\theta) \pi(\theta)}{p(X)}.	
 \label{eq:posterior_bayes}
\end{equation} 

The posterior distribution contains all the information about the parameters given the observed data and therefore it is of high importance from a Bayesian viewpoint.  Although in principle we are interested in deriving its probability density function explicitly, the calculation of the normalising constant in the denominator of \eqref{eq:posterior_bayes} often makes such a task difficult. Traditionally, this has been a severe obstacle in Bayesian computation, especially when $\theta$ is high-dimensional and the constant is analytically intractable. 

Nevertheless, the last three decades have seen the development of several computational methods which enabled \emph{sampling-based} Bayesian inference; that is to be able to draw samples from $p(\theta|X)$ without the need of calculating $p(X)$. For example, Markov Chain Monte Carlo \cite{GamLop06} methods provide such a tool to sample from the posterior distribution and obtain sample estimates quantities of interest (e.g. densities, posterior moments etc.), thereby performing the integration implicitly.


When the likelihood can be either difficult or costly to evaluate it is practically infeasible to use MCMC or even to perform maximum likelihood inference.   However, provided it is possible to simulate from a model, then `implicit' methods such as Approximate Bayesian Computation (ABC) enable 
inference without having to calculate the likelihood. These methods were originally developed for applications in population genetics \cite{Pritchard}  and human demographics \cite{Beaumont}, but are now being used in a wide range of fields including epidemiology, biology and finance, to name a few; see \cite{Marin} and the references therein.

Intuitively, ABC methods involve simulating data from the model using various parameter values and making inference based on which parameter values produced realisations that are `close' to the observed data. It is easy to show that algorithm below generates {\em exact} samples from the Bayesian posterior density
$p(\theta \vert X)$ which is proportional to $p(X \vert \theta) \pi(\theta)$:
\begin{algorithm}[H]
{\bf 1}: Sample $\theta^*$ from $p(\theta)$. \\
{\bf 2}: Simulate dataset ${X^*}$ from the model using parameters $\theta^*$. \\
{\bf 3}: Accept $\theta^*$ if $X^* =X$, otherwise reject. \\
{\bf 4}: Repeat.
\caption{\newline Exact Bayesian Computation (EBC)}
\label{alg:EBC}
\end{algorithm}

Clearly, the above algorithm is of practical use only  if $\mathcal{X}$ is a discrete space, since otherwise the acceptance probability in Step 3 is zero. For continuous
distributions, or discrete ones in which the acceptance probability in step 3 is unacceptably low, \cite{Pritchard} suggests the following algorithm:

\begin{algorithm}[H]
  \vspace{0.1cm}
{\bf 1}: Sample $\phi^*$ from $\pi(\phi)$. \\
{\bf 2}: Simulate dataset ${X^*}$ from the model using parameters $\phi^*$. \\
{\bf 3}': Accept $\phi^*$ if $d\big(s(x),s(x^*)\big) \leq \varepsilon$, otherwise
reject.\\
{\bf 4}: Repeat.
\caption{\newline Approximate Bayesian Computation (ABC)}
\label{alg:ABC}
\end{algorithm}
\noindent Here $d(\cdot,\cdot)$ is a distance function or metric, usually taken to be a type of Euclidian distance, $s(\cdot)$ is a function of the data, and $\varepsilon$ is a tolerance.  Note that $s(\cdot)$ can be the identity function but in practice, to give tolerable acceptance rate, it is usually taken to be a lower-dimensional vector comprising some summary statistics that characterise key aspects of the data.

The output of the ABC algorithm is a sample from the ABC posterior density 
$$
\tilde{p}(\theta |x) =  p(\theta\vert x, d \big(s(x),s(x^*)\big)\leq\varepsilon).
$$ 
If $s(\cdot)$ is a sufficient statistic, then the ABC posterior density converges to $p(\theta \vert x)$ as $\varepsilon \rightarrow 0$ \cite{Marin}.  However, in practice it is often difficult to find an $s(\cdot)$ which is sufficient. Hence ABC requires a careful choice of $s(\cdot)$ and $\varepsilon$ to make the acceptance rate tolerably large, at the same time as trying not to make the ABC posterior too different from the true posterior $p(\theta\vert x)$. 

In section \ref{sec.abc.maser} we will apply the ABC procedure for estimating the Rabi angle of the atom maser. For this, we will first discuss different choices of summary statistics and investigate to what extent they enable us to recover the {\em true} posterior distribution. We will then show that applying ABC with a combination of the statistics gives a posterior which is comparable with the performance of the MLE on the full data.

\section{Results}
\label{sec.results}

We now return to the atom maser and apply the general frequentist and Bayesian methods described in the provious section to the problem of estimating the key dynamical parameter of the system, the Rabi angle $\phi$. In \cite {Catana} we have shown that the counts statistics $\Lambda_1(t),\Lambda_2(t)$ can be used to estimate $\phi$ and we provided explicit expressions of the corresponding (asymptotic) Fisher informations  per unit of time. Here we extend this investigation in two complementary directions: i) we study the properties of the maximum likelihood estimator (MLE) for $\phi$ and the corresponding Fisher Information when the {\em full} atom detection record is taken into account, and ii) we illustrate how a suitable ABC algorithm enables to infer $\phi$ in a likelihood-free fashion, with a similar performance to that of the full data MLE. 

\subsection{Likelihood based approaches to inference for the atom maser}
\label{sec.likelihood.atom.maser}

We will first consider the problem of identifying the unknown parameter $\phi$ in the full environment monitoring scenario (both atoms and emitted/absorbed photons are observed). Based on the description of the measurement processes outlined in section \ref{sec.simulations}, one can compute the likelihood function for a given measurement record and apply statistical techniques to investigate the asymptotic behaviour of the MLE \cite{Catana}. Although we are not in an i.i.d. setting, the process consisting of the measurement record \emph{together} with the conditional cavity state is Markovian and ergodic, and the asymptotic normality results described in section \ref{sec.frequentist} apply to this set-up. In particular, the Fisher information up to time $t$ increases linearly in time in the sense that 
$$
\lim_{t\to\infty} \frac{I(t)}{t} = I \neq 0
$$
and the mean square error of the MLE scales as $1/(I \cdot t)$. For simplicity we will call $I$ the Fisher information, but we should note that it is an asymptotic information per unit of time. As shown in \cite{Catana} the Fisher information associated to the counting measurements in all four channels, achieves the upper bound given by the \emph{quantum} Fisher information contained in the joint quantum state of system (cavity) and output (atoms and photon bath) and is proportional to the stationary mean photon number:  $I= I_{full}:= 4N_{ex} {\rm Tr}(\rho_{ss}N)$. This result will serve as a benchmark for assessing the power of other estimators constructed from the atoms detection process which \emph{is} accessible in experiments.

We switch now to the second scenario where the data consists solely of the atoms measurement record. Since we average over photon emission and absorption events, the measurement process is \emph{not} Markovian, but falls in the class of hidden Markov processes, where the observations (atom detections) are certain stochastic functionals of an unobserved Markov process (the cavity). 

Below we analyse two likelihood-based sub-scenarios. In the first one \cite{Catana}, all time correlations in the detection record are ignored and we only consider the \emph{total number of counts} of a certain type up to time $t$, for which the Fisher information can be computed explicitely. In the second sub-scenario we apply the maximum likelihood method to the full atom counts record, and estimate the corresponding Fisher information.

\subsubsection{Estimation based on total number of detections statistics}
\label{sec.total.counts}

Let $\Lambda(t) = (\Lambda_{1}(t), \Lambda_{2}(t))$ be the two total number of counts up to time $t$, as defined in section \ref{sec.simulations}. 
By applying the Markov property one can compute the characteristic function of 
$\Lambda(t)$ and prove that the processes are (jointly and locally) asymptotically normal \cite{Catana}. 
More precisely, if $\phi= \phi_0+ u/\sqrt{t}$ then the properly rescaled count statistics converge to a Gaussian distribution 
\begin{equation}\label{eq.asymptotic.normality.Lambda}
\frac{1}{\sqrt{t}} \left( \Lambda(t) - \mathbb{E}_{\phi_{0}}(\Lambda(t)) \right) \overset{\mathcal{L}}{\longrightarrow} N\left(\mu u, V\right),
\end{equation}
where both $\mu=\mu(\phi_0, N_{ex})$ and $V=V(\phi_0, N_{ex})$ have explicit expressions in terms of certain Lindblad type generators on the cavity space. 
Note that since $\phi$ is unknown,  $\Lambda(t)$ is not `centred' by subtracting its expectation $\mathbb{E}_{\phi}(\Lambda(t))$  as customary in the Central Limit Theorem, but with the expectation at the fixed parameter $\phi_0$. Consequently, the Gaussian limit has a non-zero mean which is proportional to the local parameter $u$. 

\begin{figure}[H]
\centering
  \scalebox{0.7}{\includegraphics[width=0.8\textwidth]{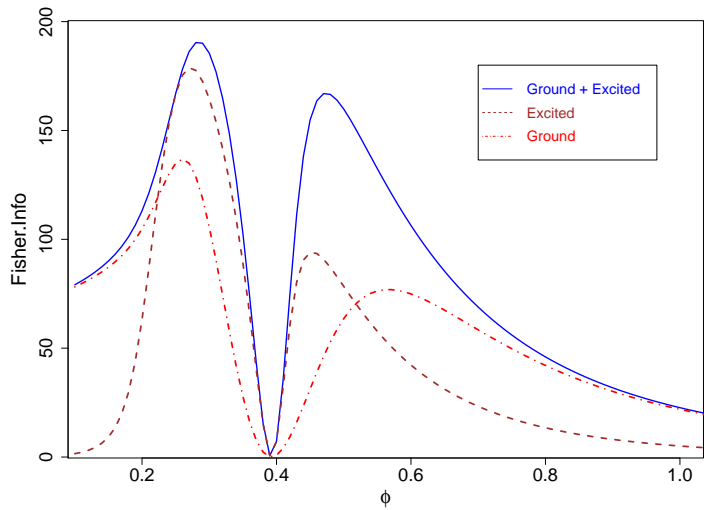}}
 \caption{Fisher informations (per unit of time) for different count statistics as function of $\phi$, for 
 $N_{ex}=16$: total number 
 of ground state atoms  (dash-dotted red curve), total number of excited state atoms (dashed brown curve), optimal Fisher information (continuous blue curve). }
    \label{fig:counts}
\end{figure}

From the Gaussian limit model we can easily obtain the Fisher information for different combinations of the count statistics. In particular the Fisher informations for the components $\Lambda_1$ and $\Lambda_2$ are
$$
I_1 = \frac{\mu_1^2}{V_{11}}, \qquad I_e = \frac{\mu_2^2}{V_{22}}
$$
and their dependence on $\phi_0$ is illustrated in Figure \ref{fig:counts} by the blue and red lines. 
By optimising over the linear combinations $a_1\Lambda_1(t)+ a_2\Lambda_2(t)$ we obtain the maximum Fisher information that can be 'extracted' from the total counts statistics
\begin{equation}\label{eq.I.opt.total.counts}
I^*= \max_{a^t a=1}\frac{(a^t \mu)^2 }{ a^t Va }.
\end{equation}
The graphs of the three Fisher informations as functions of $\phi$ are shown in Figure \ref{fig:counts}. We note that all informations are equal to zero at a point $\phi\approx 0.4$ where the means of $\Lambda_1, \Lambda_2$ are stationary with respect to $\phi$.

\subsubsection{Estimation based on the full likelihood} \label{flik}

\begin{figure}[H]
\centering
  \scalebox{0.6}{\includegraphics[width=0.9\textwidth]{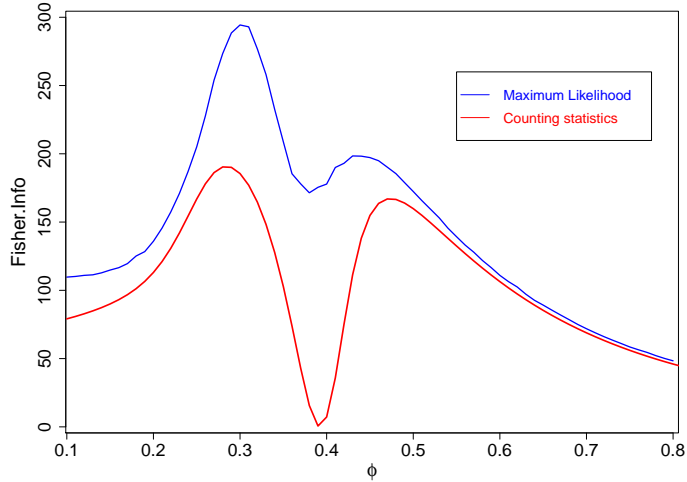}}
 \caption{The (estimated) Fisher information of the full measurement record $I_{full}$ (blue curve) versus the Fisher information of the partial likelihood of the total counts statistics $I^*$ (red curve) for $N_{ex}=16$. The former is much larger than the latter in the region around  $\phi\approx 0.4$ where $I^*\approx 0$.}
    \label{fig:mle}
   \end{figure}

We now investigate the performance of the MLE and the corresponding Fisher information in the case where the full measurement record up to time $t$ is taken into account, rather that only the total counts. 
As outlined in section \ref{sec:detection},  a detection record 
${\bf d}_{t]} := \{ (t_1, i_1), \dots ,(t_n, i_n)\}$ consists of a sequence of detection  times prior to time $t$, together with labels indicating the outcome of each atom measurement. In simulations, such a record can be generated by following the procedure described in section \ref{sec.simulations}. By construction, this is a hidden Markov process, where the underlying Markov dynamics is that of the cavity. 
The likelihood of the data ${\bf d}_{t]}$ , seen as a function of $\phi$, is  
$
p_\phi\left({\bf d}_{t]}\right) = {\rm Tr}\left(\rho\left(t~;~ {\bf d}_{t]}\right)\right),
$
where $\rho\left(t~;~ {\bf d}_{t]}\right)$ is the unnormalised cavity state defined in \eqref{eq:condstate}. For computing the MLE it is more convenient to work with the log-likelihood function
$\ell_\phi \left({\bf d}_{t]}\right):= \log p_\phi \left({\bf d}_{t]}\right)$, which can be expressed as
$$
\ell_\phi \left({\bf d}_{t]}\right) = \sum_{k=1}^n \log {\rm Tr} \left(\mathcal{J}_{i_{k}} e^{\mathcal{L}_{0}(t_{k}-t_{k-1})}\rho_{k-1}\right) + \log {\rm Tr}\left( e^{\mathcal{L}_{0}(t-t_{n}) }\rho_n\right)
$$
where $\rho_k$ denotes the conditional cavity state after the $k$'th jump
\begin{equation*}
\rho_{k}=\frac{\mathcal{J}_{i_k}  e^{\mathcal{L}_{0}(t_{k}-t_{k-1})}\rho_{k-1}}{{\rm Tr} \left(\mathcal{J}_{i_{k}} e^{\mathcal{L}_{0}(t_{k}-t_{k-1})}\rho_{k-1}\right)}.
\end{equation*}
Additionally,  since the initial cavity state can be chosen to be the (diagonal) stationary state $\rho_{ss}$, the jumps and semigroup calculations can be restricted to diagonal states (probability distributions) which are truncate to a sufficiently large photon number. With these notations, the maximum likelihood estimator is defined as 
$$
\hat{\phi}_t := \arg\max_{\phi^\prime} \ell_{\phi^\prime} \left({\bf d}_{t]}\right). 
$$
Since the theoretical expression of the Fisher information $I_{full}$ involves a complicated optimisation \cite{Douc}, we follow standard statistical methodology and use the \emph{observed} Fisher information as an estimator of the theoretical one. The former is defined as  minus the second derivative of the log-likelihood function evaluated at the MLE 
\begin{equation}
\hat{I}_{full}=
-\left.\frac{\partial^{2}\ell_\phi  \left({\bf d}_{t]}\right) }{ \partial \phi ^{2}} \right|_{\phi=\hat{\phi}_t}.
\end{equation}
Figure \ref{fig:mle} shows the average of the observed Fisher information over 150 trajectories 
(blue curve) versus the optimal Fisher information $I^*$ contained in the total count statistics, defined in \eqref{eq.I.opt.total.counts} . As expected, the former is always larger than the latter; more remarkably, the figure shows that the 
full measurement data is much more informative that the total counts statistics in the region around the 
$\phi\approx0.4$ where the mean photon number attains a local maximum and  $I^*$ is equal to zero.     

\subsection{ABC for the Atom-Maser}
\label{sec.abc.maser}


In this section we apply the ABC procedure to the  experimental data consisting of a set of detection events 
${\bf d}_{t]} := \{ (t_1, i_1), \dots ,(t_n, i_n)\}$, and different combinations from a total of 7 summary statistics. The statistics fall into four categories: i)  total number of ground and excited state atoms; ii) waiting time statistics for both ground and excited state atoms; iii) average number of the consecutive detections of the same type; iv) a statistic of the local density of detected ground state atoms in the vicinity of detected excited atoms. For a better physical intuition we will give a brief description of each statistic, but we emphasize that no knowledge of their theoretical properties is required in  applying the ABC estimation procedure.

Furthermore, we employ different distance metrics for each type of summary statistic. For the total numbers of atoms we use the Euclidean distance, while for the average number of consecutive detections the distance is defined as the absolute value of the difference divided  by the average number of consecutive detections in the experimental data. When using the local density and the waiting time statistics we essentially need a metric which measures the distance between two random samples. The Kolmogorov--Smirnov (KS) distance is often used to test whether two underlying one-dimensional probability distributions differ. Therefore, we employed the statistic which is used for a Kolmogorov--Smirnov test as our distance metric and is defined as follows. 

Let $X$ and $Y$ be two random variables with distributions $\mathbb{P}_X$ and  $\mathbb{P}_Y$ respectively.  Denote their cumulative distribution functions (CDF) by $F(x)=\mathbb{P}_X(X \leq x)$ and  $G(x)=\mathbb{P}_Y(Y \leq x)$ respectively. Given $n$ identically and independently distributed observations ${\bm X} = (X_1, \ldots, X_n)$ and ${\bm Y} = (Y_1, \ldots, Y_n)$ from $\mathbb{P}_X$ and respectively $\mathbb{P}_Y$, we denote by $F_n(x)$ and $G_n(x)$ denote the empirical cumulative distribution functions, i.e
$$ 
F_n(x) = (1/n) \sum_{i=1}^{n} {\bm 1}_{\{X_i \leq x\}}, \qquad G_n(y) = (1/n) \sum_{i=1}^{n} {\bm 1}_{\{Y_i \leq y\}}.
$$ 
The KS distance between the two samples is defined as 
$$
KS({\bm X},{\bm Y}) = \sup_{x} \left(F_n(x)-G_n(x)\right ).
$$

The ABC algorithm we use is a slight variation on the general one described previously. For a given set of experimental data corresponding to a value of the unknown parameter $\phi_{true}$ we perform a number $n = 2\times 10^6 $ simulations  for trial parameters $\phi_{trial} \in \left[ 0.1, 1.5\right ]$ drawn with uniform probability. For each simulation we store the value of the parameter and the values of each of the 7 distances between the corresponding statistics of the simulated data and the statistics of the experimental data. Let $P=\{\phi_{trial} ^i, i= 1,..., n\}$ be the set of all the simulated parameters and let $D^j=\{ d_i^j, i=1,..., n\}$ be the corresponding set of distances for statistic $j \in \{1,...,7\}$. Let $D_{m}^j$ be the subset of $D^j$ that contains the smallest $5\%$ (corresponding to a small $\epsilon$) of the elements in $D_j$.  Let $P_m^j=\{\phi_{trial} ^i | d_i^j \in D_m^j \} $ be the set of parameters corresponding to the distances contained in $D_m^j$  i.e. the values of the trial 
parameter which minimize the distance between the statistic $j$ of the trial data and the experimental data. Finally, for statistic $j$, the set of accepted parameters  $P_m^j$ is used to build the corresponding posterior distribution. 

In order to improve the resulting posterior distribution of the unknown parameter we also consider  combinations of two or more statistics. This boils down to finding the parameters  which minimise the distances corresponding to \emph{all} the chosen statistics. For example, when two statistics $j$ and $k$ are considered together, the resulting posterior distribution is built from the set $P_m^{jk}=P_m^j \cap P_m^k$ and is narrower than each individual posterior distribution. Following this reasoning the best posterior distribution is obtained from the intersection of all $7$ sets $P_m^j$ and it is shown in Figure ~\ref{fig:alltogether}.

\subsubsection{Statistics of waiting times}

\begin{figure}[H]
  \hfill
    \begin{minipage}[t]{.95\textwidth}
    \begin{center}  
  \scalebox{0.65}{\includegraphics[width=0.99\textwidth]{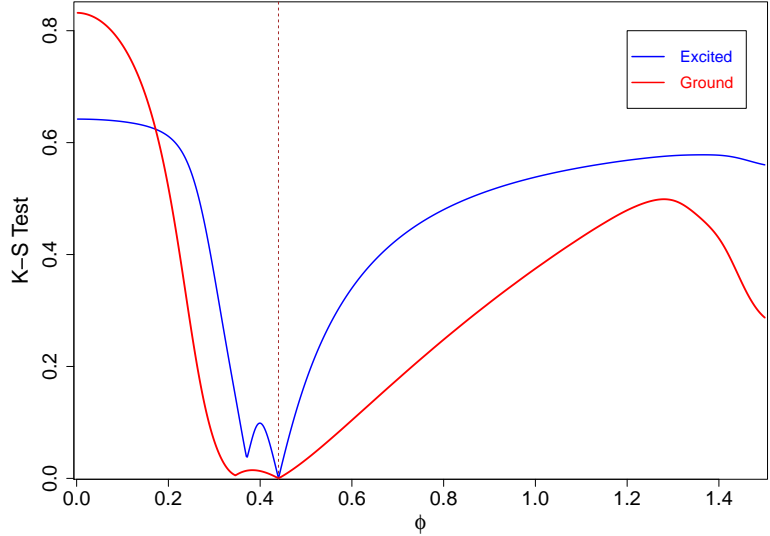}}
    \end{center}
  \end{minipage}

  \hfill
 \caption{The Kolmogorov-Smirnov distances between CDFs of the theoretical waiting time distributions (red: ground state, blue: excited state) plotted as a function of  $\phi$ at fixed $\phi_0=0.44$ (brown dotted line). }
\label{fig:kstheory}
\end{figure}

We will call a waiting time, the time between two successive detections of atoms in the \emph{same} (ground or excited) state. In the stationary regime, the state of the cavity immediately after a jump of type $i$ is on average given by 
$\rho^{after}=J_i \rho^{ss} / {\rm Tr}(J_i \rho^{ss}) $,  where $\mathcal{J}_i(\rho)= L_i\rho L_i^* $ is the  corresponding jump operator. This can be seen as follows. Being an average state, it should be equal to the state after another jump, when averaging  over all waiting times. The probability density of the waiting time, with initial state $\rho$ is  
$
p_i(t)= - {\rm Tr} \left( \mathcal{J}_i  e^{(\mathcal{L}- \mathcal{J}_i)t }  (\rho) \right) 
$
and the state after the jump is $\rho(t) = \mathcal{J}_i  e^{(\mathcal{L}- \mathcal{J}_i)t }  (\rho) / p_i(t) $. Therefore the average over time is 
$$
\rho^\prime = \int_0^\infty p_i(t) \rho(t) dt=  \int_0^\infty  \mathcal{J}_i  e^{(\mathcal{L}- \mathcal{J}_i)t }  (\rho) dt = 
 \mathcal{J}_i  (\mathcal{L}- \mathcal{J}_i)^{-1}  (\rho) 
$$
By identifying $\rho$ and $\rho^\prime$ we obtain the solution $\rho^{after}$ defined above. The corresponding waiting time is
\begin{equation}\label{eq:click}
p_i(t)={\rm Tr}\left( \mathcal{J}_i e^{(\mathcal{L}- \mathcal{J}_i)t }\mathcal{J}_i(\rho^{ss})\right) / {\rm Tr}\left( \mathcal{J}_i(\rho^{ss})\right).
\end{equation}
%
%
%
Let $F^{(w)}_i(t;\phi)$ be the cumulative distribution function of $p_i(t)$, where we have made explicit the dependance on the unknown parameter $\phi$. This dependence can be quantified by means of the Kolmogorov-Smirnov distance
$$
KS^{(w)}_i  (\phi, \phi_0)= \sup_{t\geq 0} \left|F^{(w)}_i(t;\phi) - F^{(w)}_i(t;\phi_0)\right|
$$
which is plotted in Figure \ref{fig:kstheory} as a function of $\phi$, with $\phi_0=0.44$ and stationary initial state. In general when the KS distance is large, the parameter $\phi_0$ is easier to distinguish from $\phi$ and therefore the accuracy of the ABC procedure is higher. A notable feature of Figure \ref{fig:kstheory} is the fact that the KS distance has two local minima, one for $\phi = \phi_0$ and a slightly larger one at a value for which the stationary cavity mean photon number is roughly the same as for $\phi_0$. This behaviour is related to the vanishing of the classical Fisher information for total atom counts statistics at $\phi=0.4$  as discussed in section \ref{sec.total.counts}.

\begin{figure}[H]
  \hfill
  \begin{minipage}[t]{.45\textwidth}
    \begin{center}  
       \includegraphics[width=0.99\textwidth]{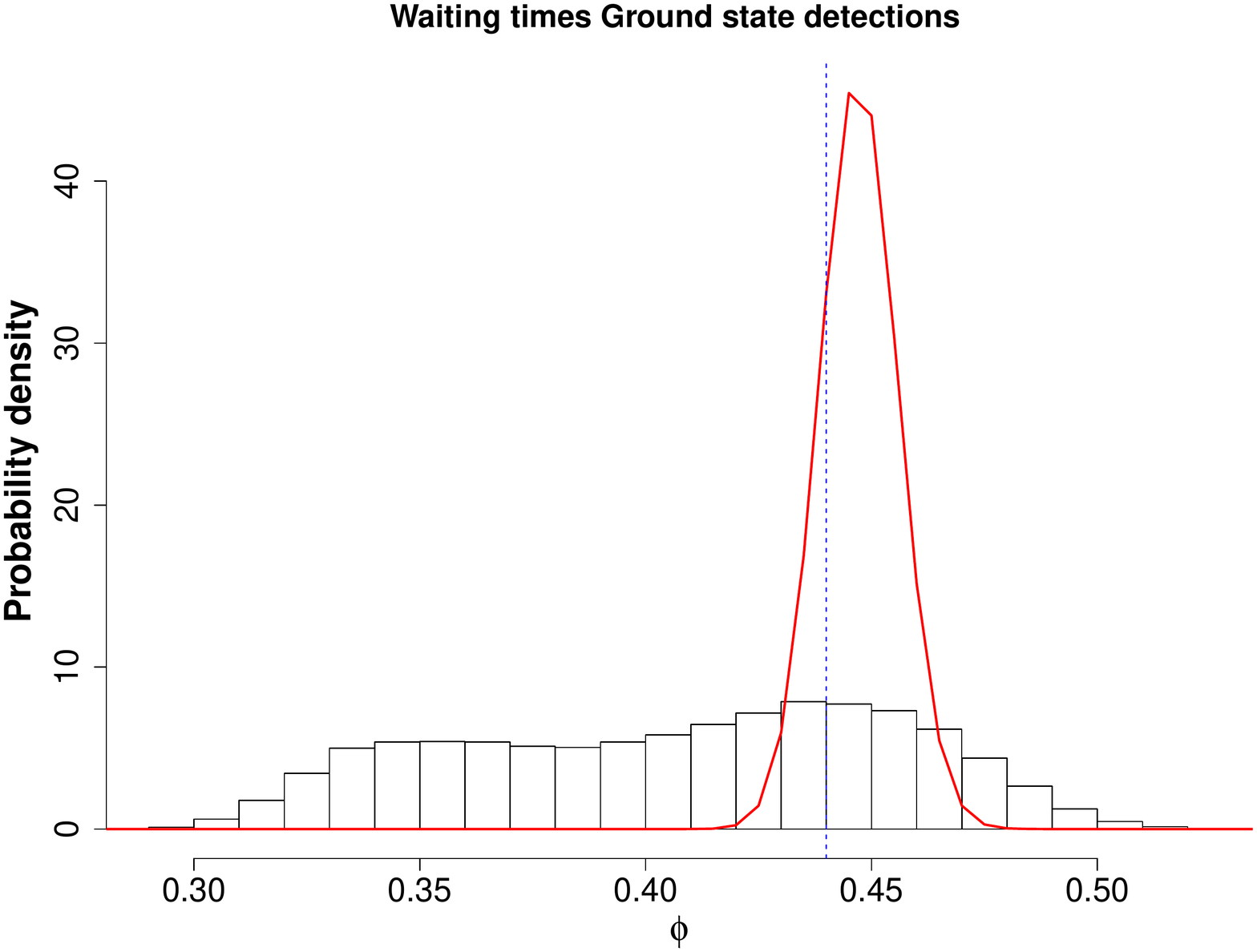}
    \end{center}
  \end{minipage}
  \hfill
  \begin{minipage}[t]{.45\textwidth}
    \begin{center}  
  \includegraphics[width=0.99\textwidth]{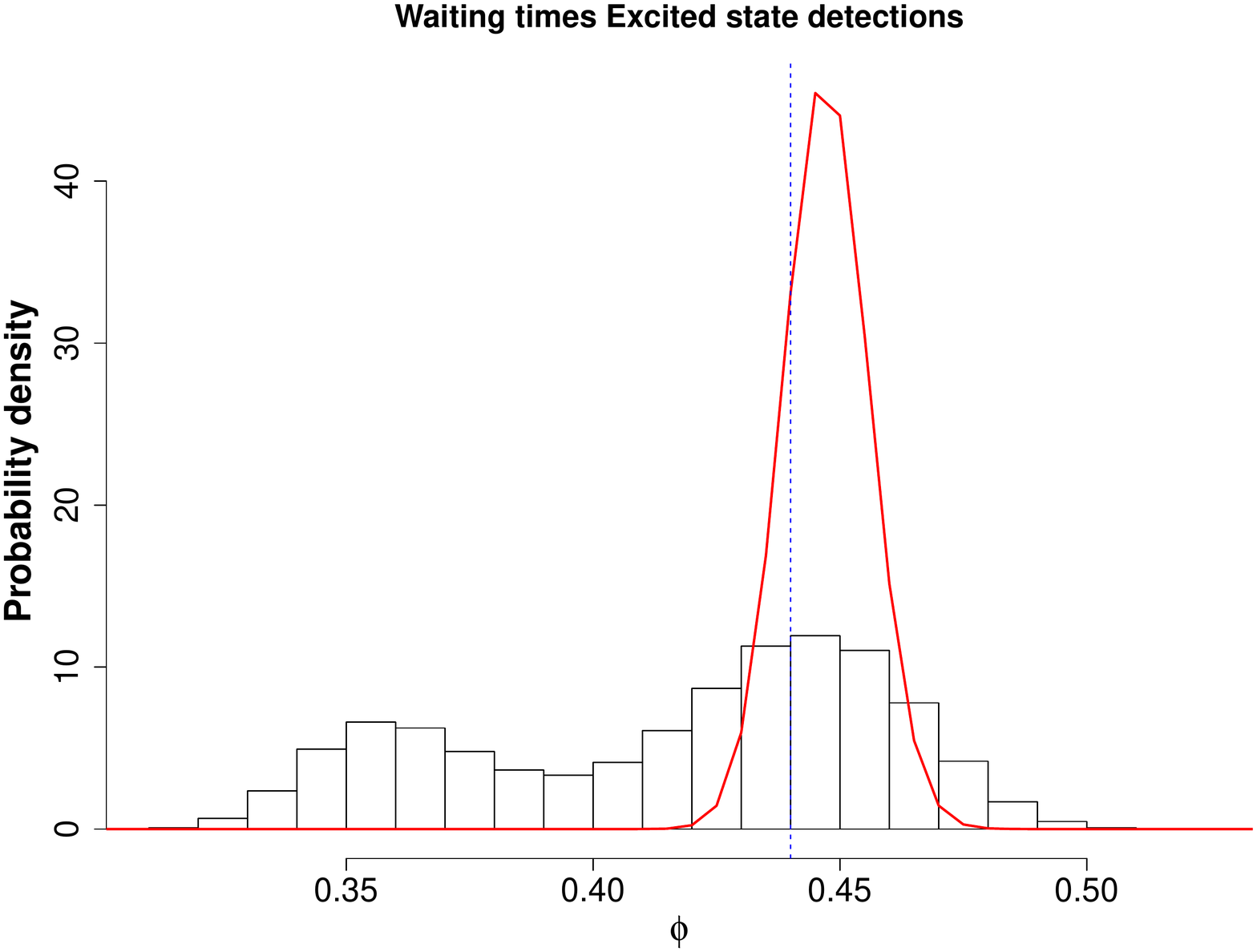}
    \end{center}
  \end{minipage}
  \hfill
 \caption{ABC posterior distribution for the waiting time statistics with $\phi_{true}=0.44$, versus the  likelihood $p_\phi\left({\bf d}_{t]}\right)$. 
 Left panel: ground state waiting times;  Right panel: excited state waiting times.}
\label{fig:wtstat}
\end{figure}

We now apply ABC for the summary statistics given by the empirical waiting time distribution, using the KS distance between such distributions as distance function. 
The `real' measurement data ${\bf d}_{t]} := \{ (t_1, i_1), \dots ,(t_n, i_n)\}$ is generated at $\phi_{true}= 0.44$ and the synthetic data is simulated for different values of $\phi$, as detailed in Algorithm 3. The result is the posterior distribution for this particular statistic, plotted  in Figure \ref{fig:wtstat} against the likelihood $p_\phi\left({\bf d}_{t]}\right)$, seen as posterior distribution of $\phi$ for a flat prior. As expected from the shape of the theoretical values of the KS test of Figure ~\ref{fig:kstheory}, the posterior distribution is broad and bimodal, and performs worse than the  distribution of the maximum likelihood for the 'experimental data' (red curve).

\subsubsection{Successive detector clicks}

Another statistic of interest is the \emph{empirical mean} of successive detections of atoms in the same (ground or excited) state, which is a measure of correlations between the emerging atoms \cite{Schieve}. Since these correlations depend on the value of $\phi$ this easily computable statistic  may constitute a suitable resource for estimation. 
The \emph{theoretical} mean in the stationary regime can be computed using the trajectories formalism \cite{Schieve}. The probability of the detected atom to be in the state  $a\in\{1,2\}$ when the detector clicks once is
$$
p(a)=\frac{{\rm Tr}\lbrace\mathcal{J}_a \rho^{ss}\rbrace}{{\rm Tr}\lbrace(\mathcal{J}_1+\mathcal{J}_2)\rho^{ss}\rbrace}.
$$
Let $p(b | a)$ the probability of having an $a$ detection at $t_0=0$ followed by a $b$ detection at any future time moment with no other detections in between. This is given by  
\begin{equation*}
p(b| a)=\int_{0}^{\infty}dt{\rm Tr}\left\{\mathcal{J}_b e^{\mathcal{L}_0 t} \mathcal{J}_a\rho^{ss}\right\}
\frac{1}{{\rm Tr}\left\{ \mathcal{J}_a\rho^{ss} \right\} }\\
=\frac{-1}{{\rm Tr}\left\{ \mathcal{J}_a\rho^{ss} \right\}}
{\rm Tr}\left\{\mathcal{J}_b\mathcal{L}_0^{-1}\mathcal{J}_a\rho^{ss}\right\}.
\end{equation*}
Therefore the probability of having a sequence of two clicks $a,b$ independent of time is given by 
\begin{equation*}
p(ba)=p(a)p(b|a)=\frac{-1}{{\rm Tr}\{(\mathcal{J}_1+\mathcal{J}_2)\rho^{ss}\}}
{\rm Tr}\left\{\mathcal{J}_b\mathcal{L}_0^{-1}\mathcal{J}_a\rho^{ss}\right\}.
\end{equation*}
In a similar manner we can compute the probability of more complex sequences of detections.
Denote by $p(ba^{n}b)$ the probability of having $n$ consecutive $a$ detections in between two $b$ detections, and by $p(a^{n})$ the probability of $n$ consecutive $a$ detections independent of previous or future detections. Then according to \cite{Schieve}
$$
\sum_{n=1}^{\infty}n p(ba^{n}b)=\frac{p(a)}{p(b)},\qquad p(a^{n})=\frac{p(b)}{p(ab)} p(ba^{n}b).
$$
The average number of successive detections of type $a$ is given by
\begin{equation*}
\langle n_{a}\rangle=\sum_{n=1}^{\infty}n p(a^n)=\frac{p(a)}{p(ab)}=\frac{1}{p(b|a)}=\frac{-{\rm Tr}\left\{ \mathcal{J}_a\rho^{ss} \right\}}
{{\rm Tr}\left\{\mathcal{J}_b\mathcal{L}_0^{-1}\mathcal{J}_a\rho^{ss}\right\}}.
\end{equation*}

\begin{figure}[H]
  \hfill
    \begin{center}  
     \scalebox{0.6} {  \includegraphics[width=0.99\textwidth]{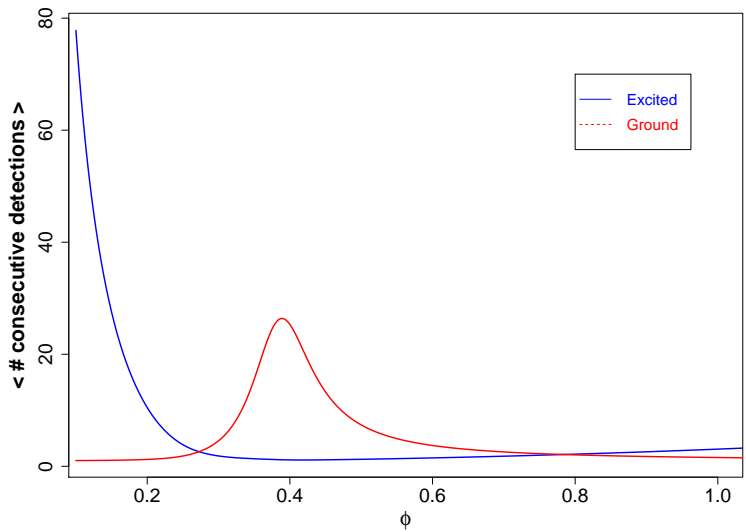}}
    \end{center}

  \hfill
 \caption{Average number of successive detections of atoms of a given type as function of $\phi$, for $N_{ex}=16$ and $\nu=0.1$, for excited state atoms (blue line) and ground state atoms (red line). }
\label{fig:cclickst}
\end{figure}

In Figure ~\ref{fig:cclickst} we plot the theoretical values of the average number of detector clicks of the same type as a function of $\phi$ for $N_{ex}=16$ and $\nu=0.1$. 
The mean number of successive ground state atoms shows a strong peak at $\phi=0.4$ and is an informative statistic around this point, although it does not differentiate between values of $\phi$ which give the same rate of production of ground state atoms. The mean number of excited state atoms is very high when the rate of these atoms is hign but varies little with 
$\phi$  beyond $0.4$.

As before, we apply the ABC procedure for the `experimental data' ${\bf d}_{t]} := \{ (t_1, i_1), \dots ,(t_n, i_n)\}$ generated at $\phi_{true}= 0.5$, and the synthetic data is simulated for different values of $\phi$, as detailed in Algorithm 3.

\begin{figure}[H]
  \hfill
  \begin{minipage}[t]{.45\textwidth}
    \begin{center}  
       \includegraphics[width=0.99\textwidth]{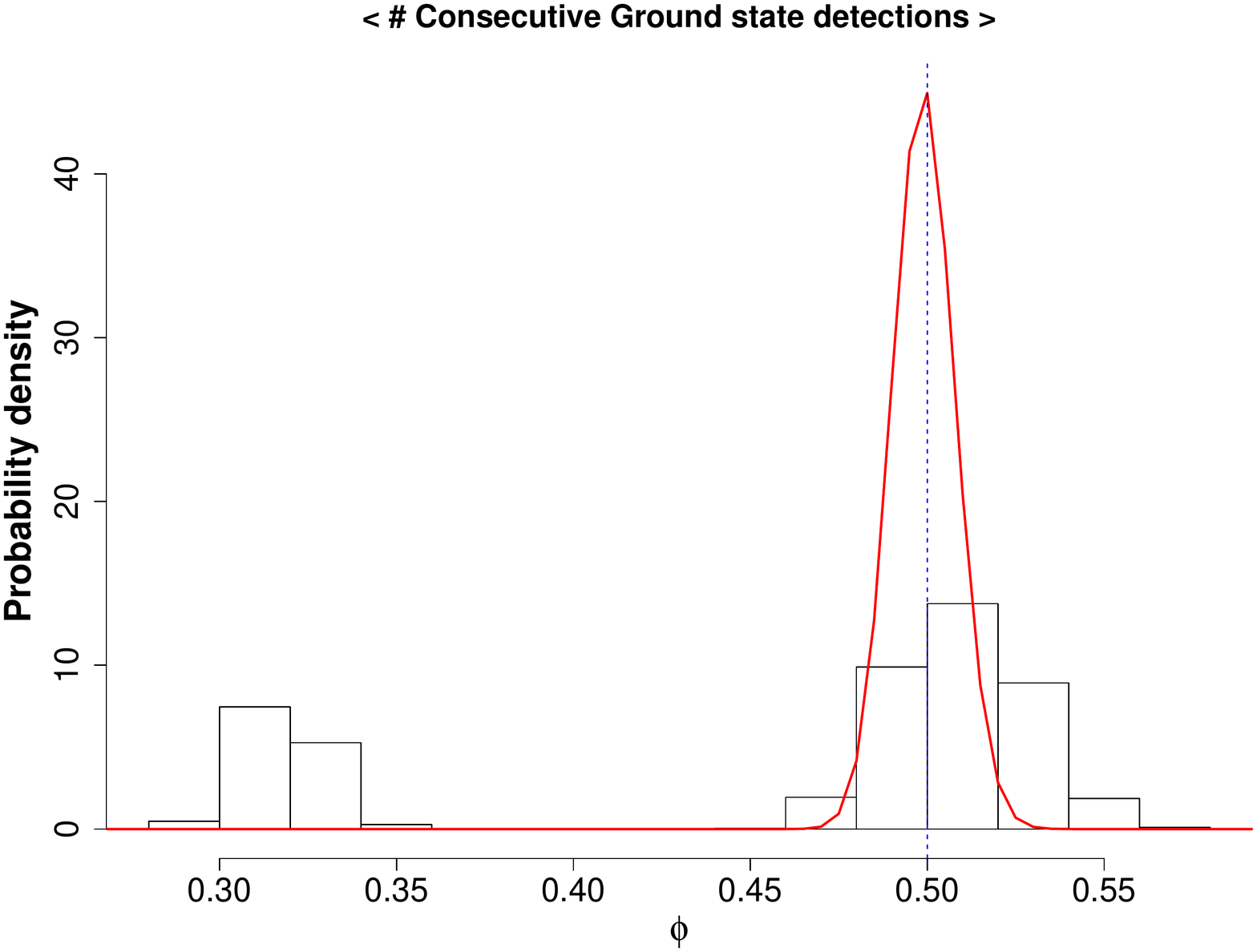}
    \end{center}
  \end{minipage}
  \hfill
  \begin{minipage}[t]{.45\textwidth}
    \begin{center}  
  \includegraphics[width=0.99\textwidth]{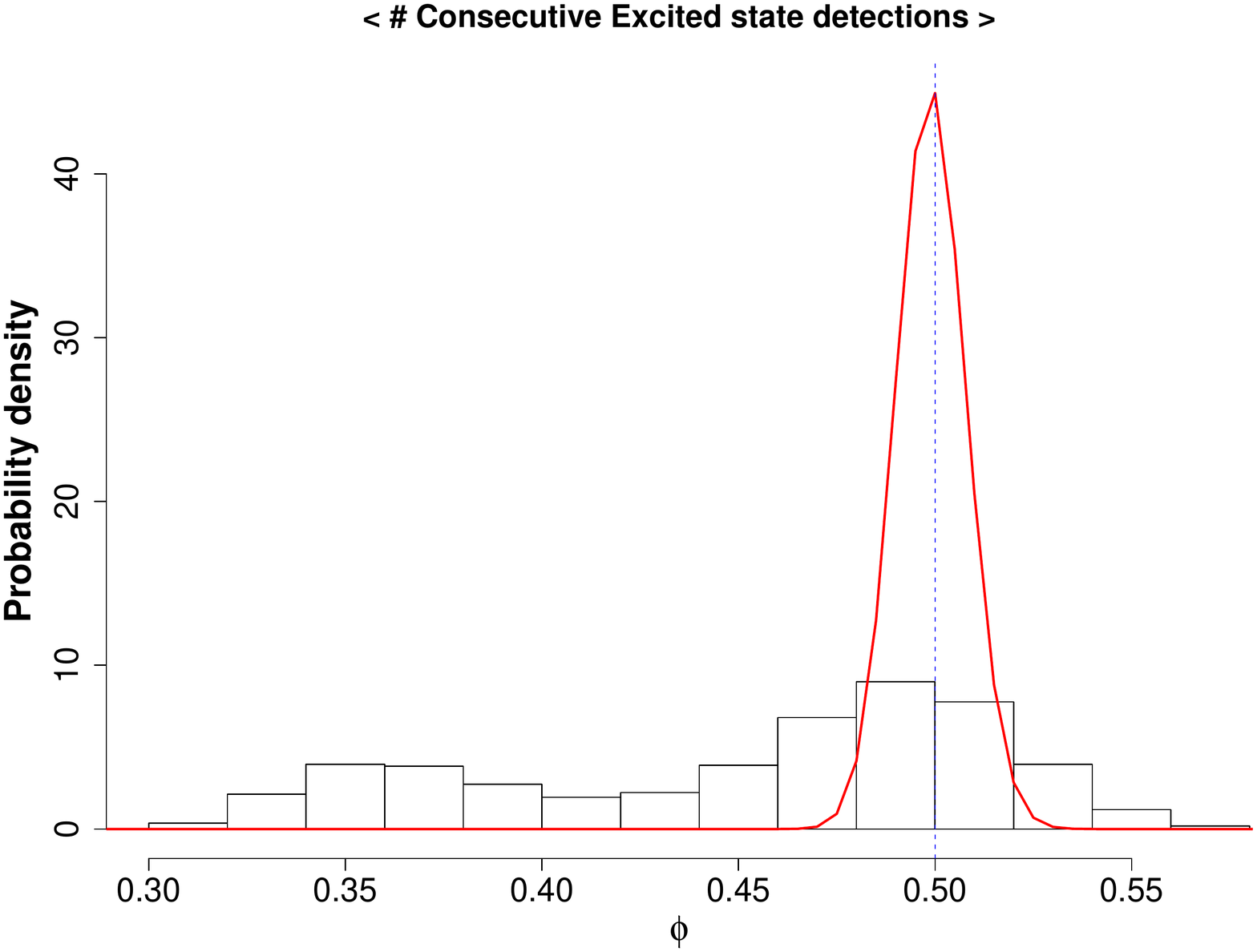}
    \end{center}
  \end{minipage}
  \hfill
\caption{ABC posterior distribution for average number of detections of the same type (histogram) at $\phi_{true}=2$,  for $N_{ex}=16$ and $\nu=0.1$, versus the likelihood of the data (red line).  Left panel: ground state atoms. Right panel: excited state atoms.}
\label{fig:consclick}
\end{figure}

For each simulation run we compute the average numbers of succesive detector clicks of the same type 
$\langle n_{1}^{sim}\rangle$ and $\langle n_{2}^{sim}\rangle$, and calculate their deviation from the averages obtained from the 
`experimental data' $\langle n_{1}^{exp}\rangle$, $\langle n_{2}^{exp}\rangle$ by the relative distance
$$
d_a=\left | 1- \frac{\langle n_{a} ^{sim}\rangle}{ \langle n_{a}^{exp}\rangle} \right| , \qquad a\in\{1,2\}.
$$
Figure \ref{fig:consclick} shows the posterior distributions obtained from the ABC method for the average number of consecutive detector clicks with the distance defined above.
 In the first case, the distribution is more concentrated around the true value, but has a secondary peak at the point with the same mean.  

\subsubsection{Local density}

We now consider a more complex statistic which takes into account both the arrival times and the number of detector clicks of ground state atoms. For each excited state atom arriving at a time $t^2_i$, we count how many ground state atoms have been detected in the time interval $(t^2_i - s, t^2_i)$. The local density is this number of atoms divided by the length of the time interval $s$. For example the following set of detection events
$$
t_1^1 = 0.1, \, t_2^1=0.25,\, t_3^1=0.37, \, t_4^1 = 0.56, \, t_5^1=0.82, \, t_6^2=1.12,\,  t_7^2= 1.21, \, t_8^1=1.33,\, t_9^2=1.67
$$
has a vector of local densities $\{l_6=4, l_7=4, l_9=2\}$  for $s=1$. 

An analytic expression for the probability distribution of the local density follows from equation \eqref{eq:click}. First   the probability of having no ground state atom in the time interval $(0,t)$ is
\begin{equation*}
p_{1}^{0}(t)={\rm Tr}\left\{ e^{(\mathcal{L}-\mathcal{J}_1)t}\rho^{ss}\right\}.
\end{equation*}
We have chosen as initial state the stationary state motivated by lack of knowledge about the atoms detected before $t_0=0$. Similarly we find the probability of $n$ detections of ground state atoms  between $t_0=0$ and  $t_1=t$, moment at which an excited atom was detected,  reads
\begin{equation}{\label{eq:countt}}
p_{1}^{n}(t)=\int_{0}^{t}dt_{1}\dots\int_{t_{n-1}}^{t}dt_{n} \,{\rm Tr}\{\mathcal{J}_2 e^{(\mathcal{L}-\mathcal{J}_1)(t-t_{n})}
\mathcal{J}_1e^{(\mathcal{L}-\mathcal{J}_1)(t_{n}-t_{n-1})}\dots \mathcal{J}_1 e^{(\mathcal{L}-\mathcal{J}_1)t_{1}}\rho^{ss}\}.
\end{equation}
According to our definition, the probability of a given local density $n/t$ is $p(n/t)=\frac{p_i^{n}(t)}{t}$. However, these theoretical probabilities are difficult to compute numerically and therefore we switch to the Bayesian approach.

\begin{figure}[H]
\centering
  \scalebox{0.5}{\includegraphics[width=0.9\textwidth]{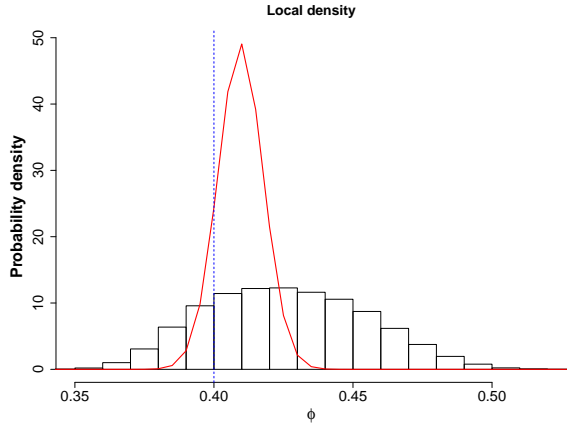}}

 \caption{ABC posterior distribution from local density of ground state atoms at $\phi_{true}=0.4$ (histogram), versus the likelihood of the real data.}
    \label{fig:locden}
   
\end{figure}

When applying ABC for this particular statistic we use as distance the Kolmogorov-Smirnov test. 
In Figure ~\ref{fig:locden} we show the posterior distribution for this statistic for $\phi_{true}=0.4$.

\subsubsection{Statistics of total counts}

The distribution of the total number of detections of atoms in state $a \in \{1,2\}$ is similar to \eqref{eq:countt} 
\begin{equation}{\label{eq:totcount}}
p_{a}^{n}(t)=\int_{0}^{t}dt_{1}\dots\int_{t_{n-1}}^{t}dt_{n}{\rm Tr}\{ e^{(\mathcal{L}-\mathcal{J}_a)(t-t_{n})}
\mathcal{J}_a e^{(\mathcal{L}-\mathcal{J}_a)(t_{n}-t_{n-1})}\dots \mathcal{J}_a e^{(\mathcal{L}-\mathcal{J}_a)t_{1}}\rho^{ss}\}.
\end{equation}
Even though these distributions can be cumbersome to compute numerically, it is easy to find the average values of ground and excited atoms rates from energy balance considerations as $\langle N\rangle-\nu$ and respectively $N_{ex}-\langle N\rangle + \nu$, where $\langle N\rangle$ is the average number of photons in the cavity in the stationary state shown in Figure \ref{fig.stationarystate}.

\begin{figure}[H]
  \hfill
  \begin{minipage}[t]{.45\textwidth}
    \begin{center}  
       \includegraphics[width=0.90\textwidth]{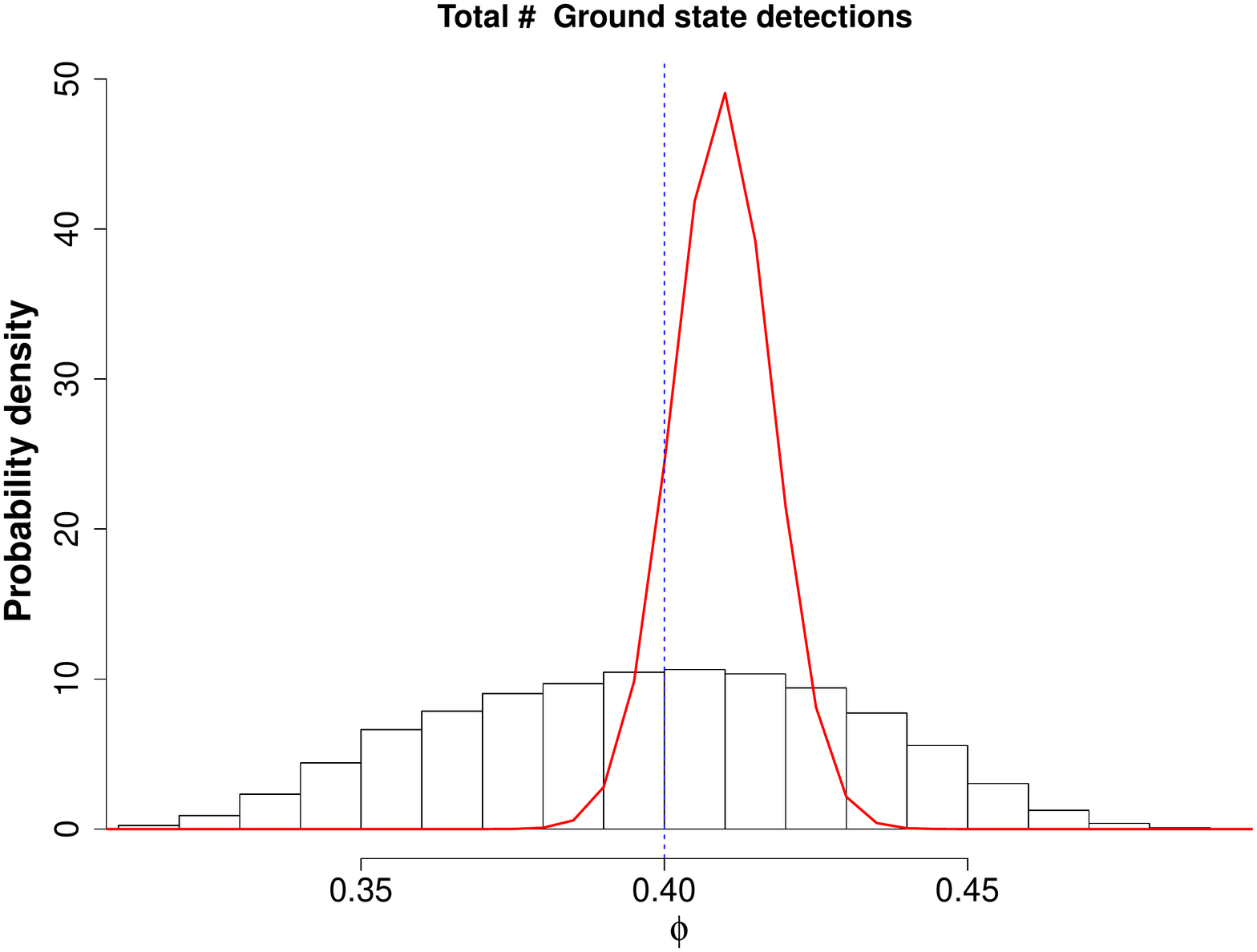}
    \end{center}
  \end{minipage}
  \hfill
  \begin{minipage}[t]{.45\textwidth}
    \begin{center}  
  \includegraphics[width=0.90\textwidth]{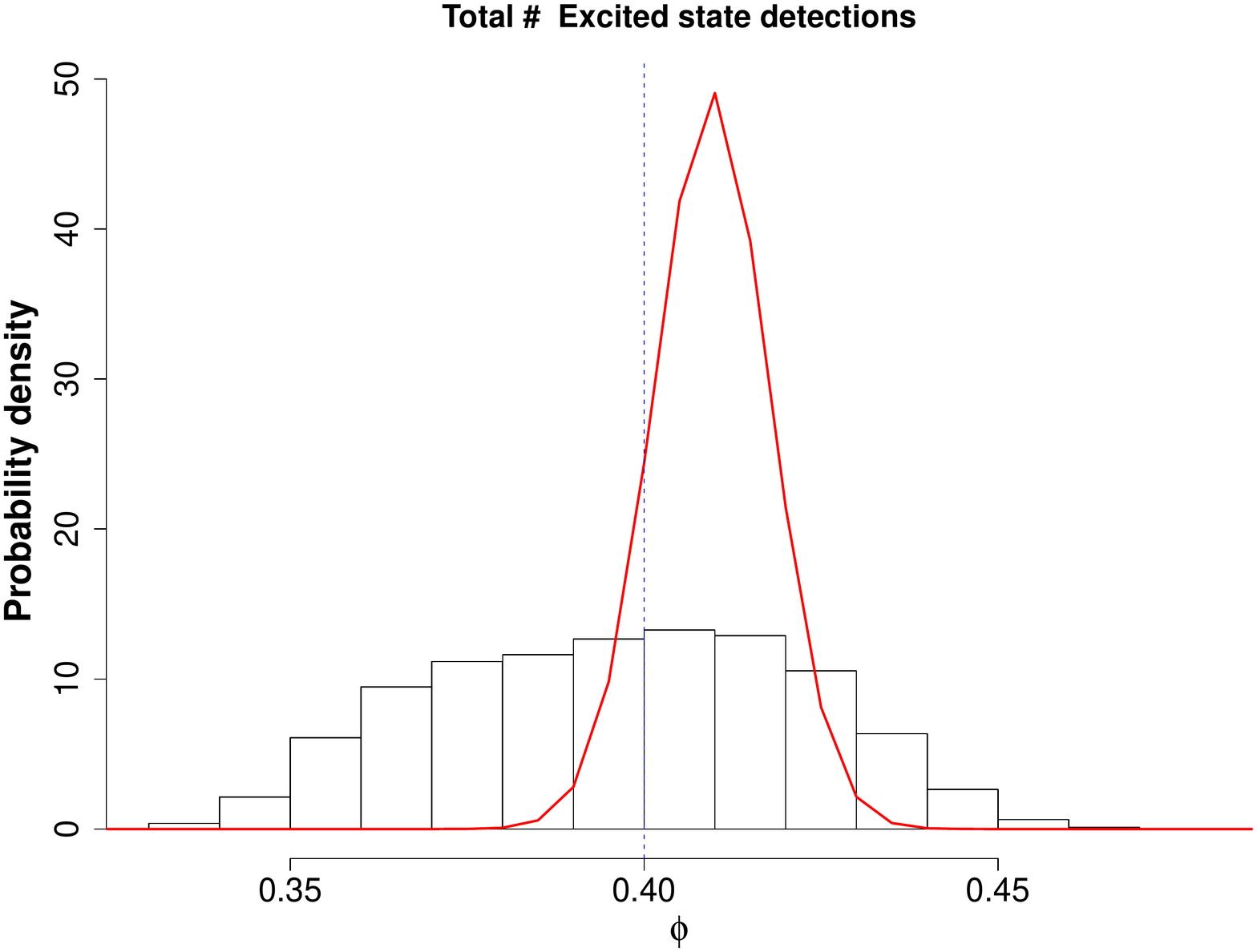}
    \end{center}
  \end{minipage}
  \hfill
\caption{ABC posterior distribution for total counts (histogram) at $\phi_{true}=0.40$, versus the likelihood of the data (red line).}
\label{fig:totcount}
\end{figure}

We apply the ABC procedure and compare the total number of atom counts in the trial simulation, with the corresponding experimental values. The resulting posterior distributions are shown in  Figure ~\ref{fig:totcount},  for $\phi_{true}=0.4$.

Close to the value $\phi=0.4$ the rate of emerging ground state atoms is at its maximum and therefore, locally in the space of parameters, the asymptotic Fisher information per unit of time, for the statistic of total counts vanishes as we have shown in the previous section and in  Figure \ref{fig:counts}. This also leads to a broad posterior distribution for the ABC method as seen in  Figure ~\ref{fig:totcount}.

\subsubsection {Combined statistics }

\begin{figure}[H]
  \hfill
  \begin{minipage}[t]{.45\textwidth}
    \begin{center}  
       \includegraphics[width=1\textwidth]{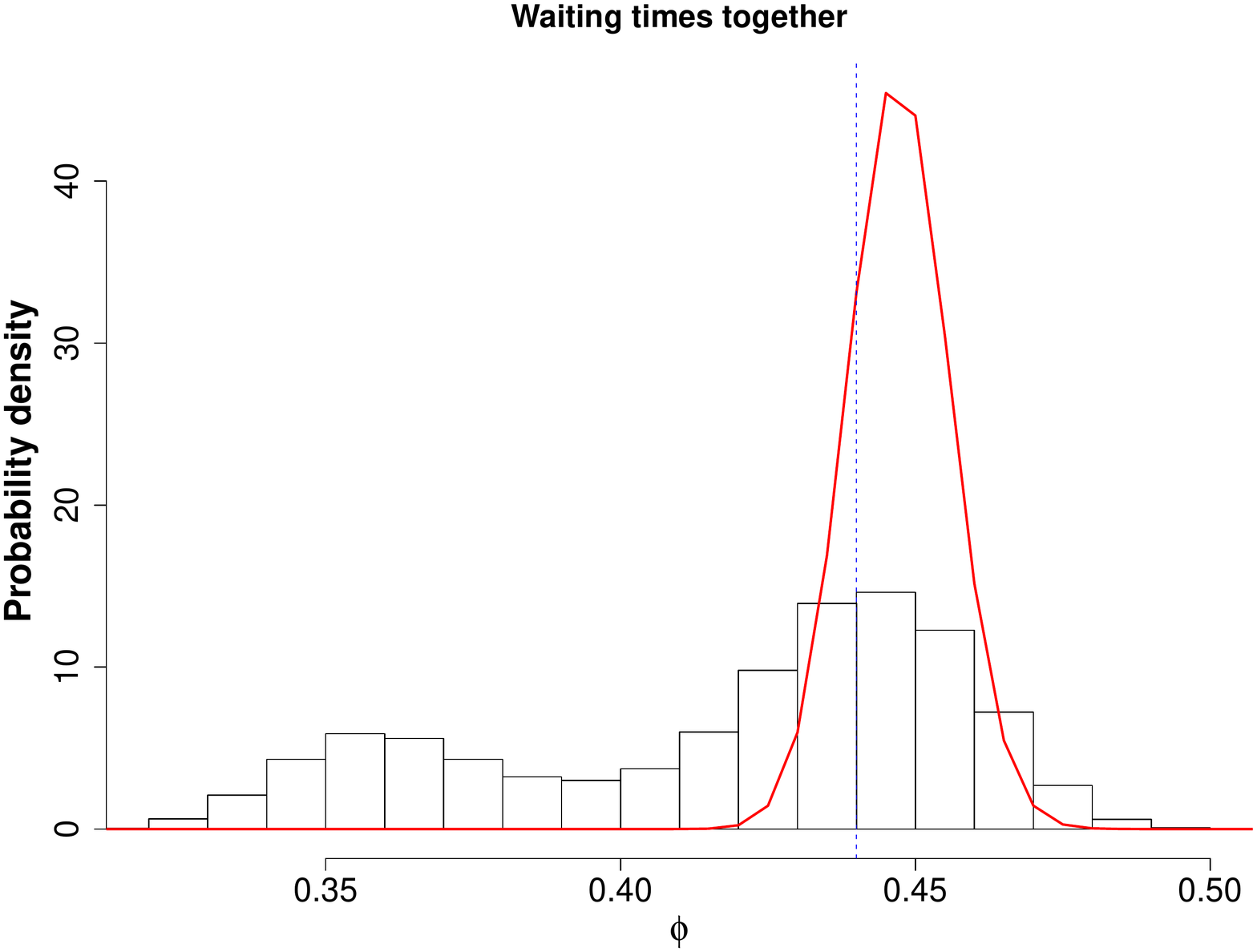}
    \end{center}
  \end{minipage}
  \hfill
  \begin{minipage}[t]{.45\textwidth}
    \begin{center}  
  \includegraphics[width=1\textwidth]{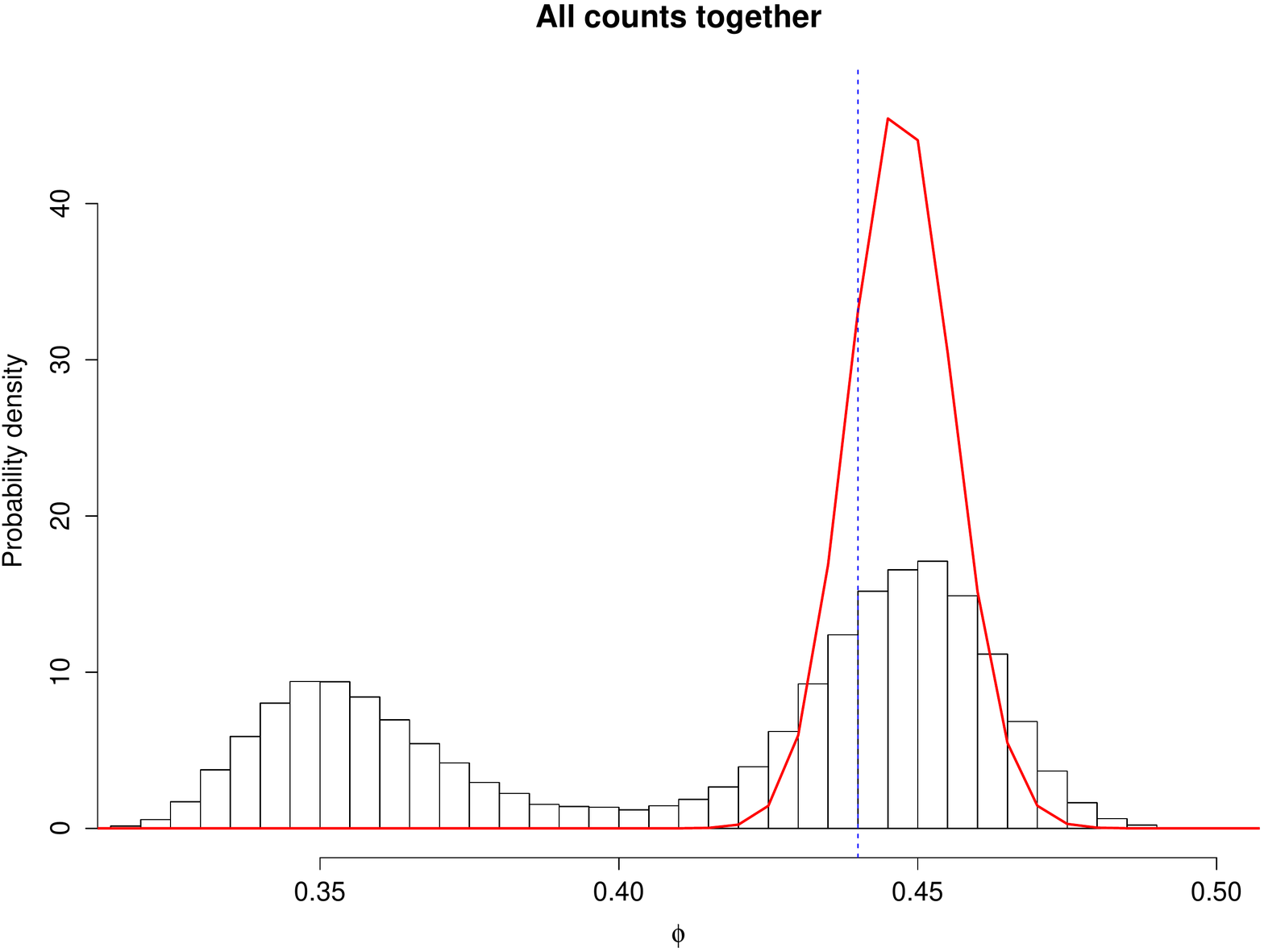}
    \end{center}
  \end{minipage}
  \hfill
   \begin{center}  
  \scalebox{0.5}{\includegraphics[width=0.99\textwidth]{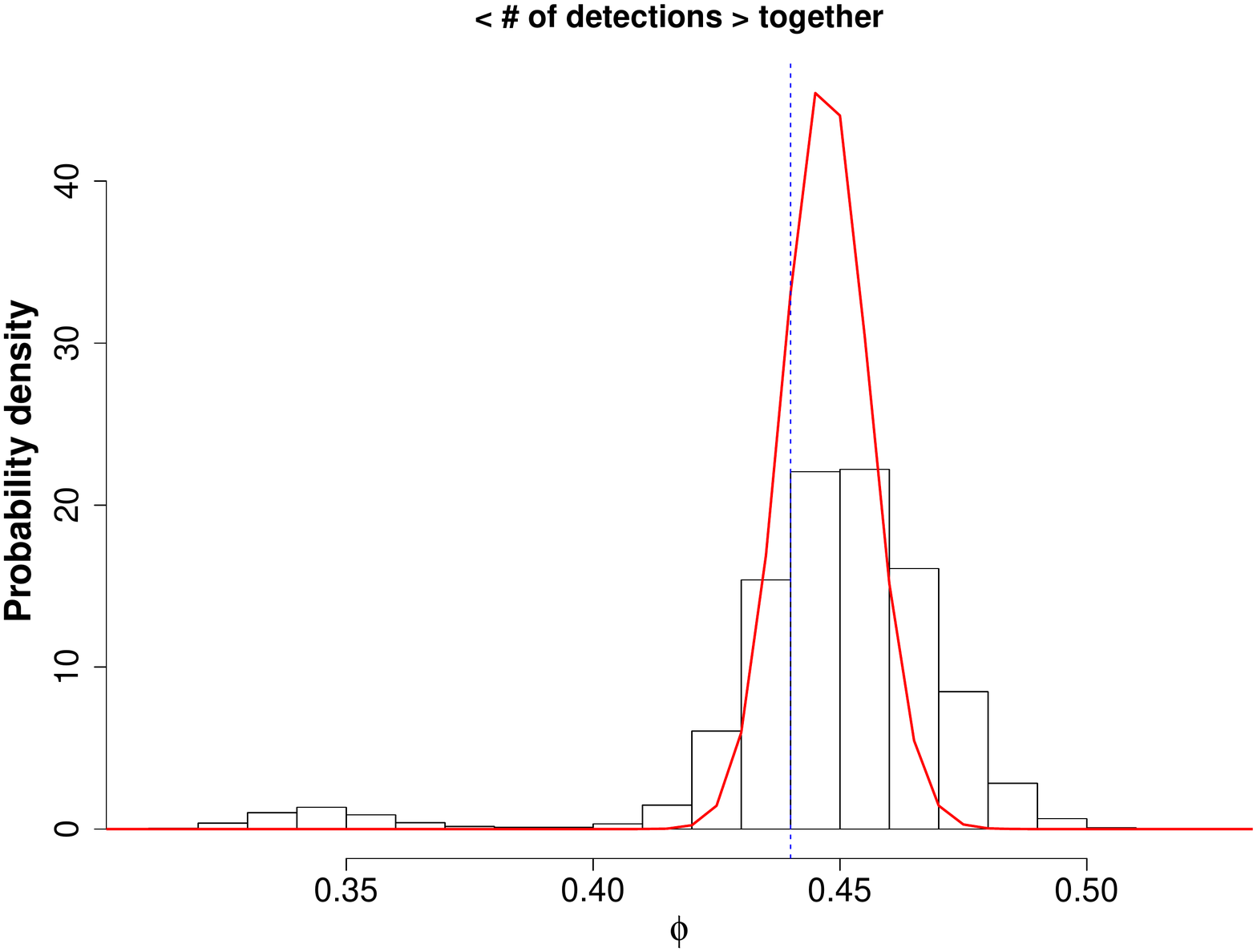}}
    \end{center}
\caption{ABC posterior distribution for combined sets of statistics (histogram) at $\phi_{true}=0.44$, versus the likelihood of the data (red line). Top left: waiting times statistics; top right: counts statistics; Bottom: both waiting times and total counts together.}
\label{fig:waitclick}
\end{figure}

As we have see, each individual statistic has a limited statistical power, and the corresponding posterior distribution is significantly broader than the likelihood. Putting two or more of the statistics together gives more information about the true parameter and improves the posterior distribution. Practically, this  boils down to taking the intersection of the sets of parameters that minimize the corresponding tests as discussed in section ~\ref{sec.abc.maser}.  In Figure ~\ref{fig:waitclick} we plot the posterior for pairs of statistics taken together and remark that the distribution gets closer to the exact posterior. The improvement  in the posterior distribution varies with the choices of statistics to be taken together. For instance, the waiting times statistics and the total number of detections look similar as they are both influenced by the rate of atoms in a given state while the statistics of the local density and average numbers of consecutive detections of the same type are quite different. 

\begin{figure}[H]
\centering
  \scalebox{0.6}{\includegraphics[width=0.9\textwidth]{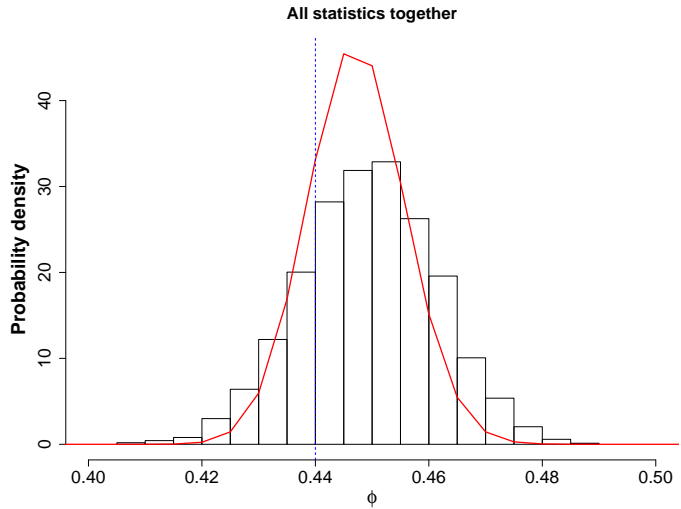}}

 \caption{ABC posterior distribution for all statistics combined (histogram) at $\phi_{true}=0.40$, versus the likelihood of the data (red line). }
    \label{fig:alltogether}
   
\end{figure}

In the end the most informative ABC procedure  is to look at all the tests together, i.e. consider the posterior distribution for the intersection of all the 7 sets. This is shown as a histogram in Figure ~\ref{fig:alltogether}, and turns out to be only slightly broader than the likelihood of the real data (red line), the latter being the posterior distribution for a flat prior.

\section{Conclusion}

We have analysed two different estimation methods for the Rabi angle of the atom maser. We have implemented a maximum likelihood procedure using the likelihood function for Hidden Markov models which turns out to coincide with the quantum filter. We have estimated the corresponding Fisher information and compare it to the results we obtained in \cite{Catana} for the total counts statistics. As expected the full measurement data is more informative than the total counts process; This is most dramatically illustrated at the point of maximum mean photon number where the Fisher information for the total counts is zero. In section \ref{sec.abc.maser} we have shown how different statistics of the emerging atoms can be used for inference within the ABC method. We have seen that, even if single statistics are not very informative and lead to poor estimates, when different statistics are taken together the estimation precision improves considerably, and is comparable with the posterior distribution of the maximum likelihood. 

Likelihood-free methods such as ABC may become a useful statistical inference tool in quantum estimation problems where traditional likelihood-based methods are difficult to implement (e.g. complicated or even unknown likelihood, non-Markovian dynamics), while simulations, or real data produced by a `benchmark' system are comparatively cheap. 

The atom maser was chosen as a study model due to its tractable probabilistic properties and its interesting physical properties. An interesting future project would be apply the ABC machinery to higher dimensional system identification problems.

\section{Acknowledgements}

The authors acknowledge the use of Nottingham University's High Performance Computer for performing the simulations involved in our study. This work was supported by the EPSRC Fellowship EP/E052290/1.


\begin{thebibliography}{1}



\bibitem{Ljung} 
L. Ljung
\newblock \emph{System Identification: Theory for the User}
\newblock Prentice Hall, 1999.

\bibitem{Fujiwara}
A. Fujiwara
\newblock Quantum channel identification problem.
\newblock \emph{Phys. Rev. A}, 63:\penalty0 042304, 2001.

\bibitem{Burgarth}
D. Burgarth, K. Maruyama and F. Nori
\newblock Indirect quantum tomography of quadratic Hamiltonians
\newblock \emph{New J. of Phys.}, 13:\penalty0 013019, 2011.

\bibitem{Cole}
J. Cole, S. Schirmer, A. Greentree, C. Wellard, D. Oi and L. Hollenberg 
\newblock Identifying an experimental two-state Hamiltonian to arbitrary accuracy
\newblock \emph{Phys. Rev. A}, 71:\penalty0 062312, 2005.


\bibitem{Howard}
M.Howard,  J. Twamley, C. Wittmann, T. Gaebel, F. Jelezko and J. Wrachtrup
\newblock   Quantum process tomography and Linblad estimation of a solid-state qubit
\newblock \emph{New J. of Phys.}, 8:\penalty0 33, 2006.

\bibitem{Mabuchi&Khaneja}
H. Mabuchi and N. Khaneja.
\newblock {Principles and applications of control in quantum systems}
\newblock \emph{Int. J. Robust Nonlinear Control}, 15:\penalty0 647--667, 2005.

\bibitem{Dowling&Milburn}
J. P. Dowling  and G. J. Milburn
\newblock Quantum technology: the second quantum revolution.
\newblock \emph{Phil. Trans. R. Soc. Lond. A}, 361:\penalty0 1655--1674, 2003.


\bibitem{Mabuchi}
H. Mabuchi
\newblock Dynamical identification of open quantum systems
\newblock \emph{Quant. Semiclass. Optics}, 8:\penalty0 1103-1108, 1996.




\bibitem{Gardiner&Zoller} 
C. Gardiner and P. Zoller
\newblock \emph{Quantum Noise}
\newblock Springer, 2004.

\bibitem{Wiseman&Milburn}
H.~M. Wiseman and G.~J. Milburn.
\newblock \emph{Quantum Measurement and Control}.
\newblock Cambridge University Press, 2009.


\bibitem{Parthasarathy} 
K. R. Parthasarathy
\newblock An Introduction to Quantum Stochastic Calculus
\newblock Birkhauser, 1992.


\bibitem{Belavkin}
V. P. Belavkin
\newblock   Quantum stochastic calculus and quantum nonlinear filtering, 
\newblock \emph{Journal of Multivariate Analysis} 42:\penalty0  171�201,  1992



\bibitem{Guta}
M. Gu\c{t}\u{a}
\newblock  Fisher information and  asymptotic normality in system identification for quantum Markov chains
\newblock \emph{Phys. Rev. A}, 83:\penalty0 062324, 2011.


\bibitem{Guta&Kiukas}
M. Gu\c{t}\u{a}, J. Kiukas.
\newblock \emph{In preparation}


\bibitem{Catana}
C. Catana, M. van Horssen, M. Gu\c{t}\u{a}
\newblock Asymptotic inference in system identification for the atom maser
\newblock \emph{Phil. Trans. R. Soc. Lond. A}, 370:\penalty0 5308-5323, 2012.


\bibitem{Meschede}
D.Meschede, H. Walther, G. Muller
\newblock One-Atom Maser
\newblock \emph{Phys. Rev. Lett.}, 54:\penalty0 551--553, 1985.

\bibitem{Englert2}
 B-G. Englert, M. Loffler, O. Benson, B. Varcoe, M. Weidinger and  H.Walther
\newblock Entangled Atoms in Micromaser Physics
\newblock \emph{Fortschr. Phys.}, 46:\penalty0 897--926, 1998.

\bibitem{Jaynes}
 E.T. Jaynes, F.W. Cummings
\newblock Comparison of quantum and semiclassical radiation theories with application to the beam maser
\newblock \emph{Proc. of the IEEE}, 51:\penalty0 89--109, 1963.




\bibitem{Englert}
B-G. Englert, T.Gantsog, A. Schenzle, C. Wagner, H. Walther
\newblock One-atom maser: Phase-sensitive measurements
\newblock \emph{Phys. Rev. A}, 53:\penalty0 4386--4399, 1996.




\bibitem{Bergou}
 J. Bergou, L. Davidovich, M. Orszag, C. Benkert, M. Hillery, M.O. Scully
\newblock Role of pumping statistics in a maser and laser dynamics : Density-matrix approach
\newblock \emph{Phys. Rev. A}, 40:\penalty0 5073--5080, 1989.

\bibitem{Guerra}
 E.S. Guerra, A.Z. Khoury, L. Davidovich, N. Zagury
\newblock Role of pumping statistics in micromasers
\newblock \emph{Phys. Rev. A}, 44:\penalty0 7785--7796, 1991.

\bibitem{Herzog}
U. Herzog
\newblock Micromaser with stationary non- Poissonian pumping
\newblock \emph{Phys. Rev. A}, 52:\penalty0 602--618, 1995.




\bibitem{Englert3}
B-G. Englert, G. Morigi
\newblock Five Lectures on Dissipative Master Equations
\newblock \emph{Lecture Notes in Physics vol. 611  : Coherent evolution in noisy environments}, Chapter 6, page 55-106,  Springer 2002.





\bibitem{Meystre}
 P. Meystre, G. Rempe, H. Walther
\newblock Very-low-temperature behaviour of a micromaser
\newblock \emph{Optics Lett.}, 13:\penalty0 73--104, 1988.



\bibitem{Bouten}
 L. Bouten, M. Gu\c{t}\u{a}, H. Maassen
\newblock  Stochastic Schro\"{o}dinger equations
\newblock \emph{J. Phys. A: Math. Gen.} 37:\penalty0  3189-3209,  2004



\bibitem{Sterpi}
 H-J. Briegel, B-G. Englert, N. Sterpi, H.Walther
\newblock One-atom maser: Statistics of detector clicks
\newblock \emph{Phys. Rev. A}, 49:\penalty0 2962--2985, 1994.


\bibitem{Merlijn}
 M. van Horssen, M. Gu\c{t}\u{a}
\newblock Large deviations, Central Limit and quantum dynamical phase transitions in the atom maser
\newblock  arXiv:quant-ph/1206.4956v2, 2013.

\bibitem{Benson}
O. Benson, G. Raithel, H. Walther
\newblock Quantum Jumps of the Micromaser Field: Dynamic Behaviour Close to Phase Transition Points
\newblock \emph{Phys. Rev. Lett.}, 72:\penalty0 3506--3509, 1994.

\bibitem{Igor}
J.P. Garrahan, A.D. Armour, I. Lesanovsky
\newblock Quantum trajectory phase transitions in the micromaser
 \newblock  \emph{Phys. Rev. E}, 84:\penalty0  021115, 2011.
 


\bibitem{Young&Smith} 
G. A. Young  and R. L. Smith
\newblock \emph{Essentials of Statistical Inference}
\newblock Cambridge University Press, 2010.



\bibitem{Petrie}
T. Petrie
\newblock Probabilistic functions of finite state Markov chains
 \newblock \emph{The Annals of Math. Statistics}, 40:\penalty0 97-115, 1969.




\bibitem{Baum}
L.E. Baum, T. Petrie
\newblock Statistical inference for probabilistic functions of finite state Markov chains
 \newblock \emph{The Annals of Math. Statistics}, 37:\penalty0 1554-1563, 1966.
 
  
\bibitem{Leroux}
B.G.Leroux
\newblock Maximum-likelihood estimation for hidden Markov models
 \newblock  \emph{Stoch. Proc. and their Applications}, 40:\penalty0 127-143, 1992.

\bibitem{Douc}
R. Douc, C. Matias
\newblock Asymptotics of the Maximum Likelihood Estimator for general Hidden Markov Models
 \newblock \emph{Bernoulli}, 7:\penalty0 381-420, 2001.


\bibitem{Ryden2}
P.J. Bickel, Y. Ritov, T. Ryden
\newblock Asymptotic normality of the Maximum likelihood estimator for general hidden Markov models
 \newblock \emph{The Annals  Statistics}, 26:\penalty0 1614-1635, 1998.


\bibitem{Holevo} 
A. S. Holevo
\newblock \emph{Probabilistic and Statistical Aspects of Quantum Theory}
\newblock North-Holland, 1982.


\bibitem{Gill}
O. Barndorff-Nielsen, R. D. Gill and P. Jupp
\newblock On quantum statistical inference 
 \newblock  \emph{J. Royal Stat. Soc. B}, 65:\penalty0 775-816, 2003.
 
 \bibitem{Gill&Guta}
R. D. Gill and M. Gu\c{t}\u{a}
\newblock On Asymptotic Quantum Statistical Inference 
 \newblock  \emph{IMS Collections}, 65:\penalty0 105-127, 2012.



\bibitem{Caves}
S.Braunstein, C.Caves
\newblock  Statistical Distance and the Geometry of Quantum States
\newblock \emph{Phys. Rev. Lett.}, 72:\penalty0 3439--3443, 1994.





\bibitem{Guta&Kahn}
M. Gu\c{t}\u{a}, J. Kahn
\newblock Local asymptotic normality for qubit states
\newblock \emph{Phys. Rev. A}, 73:\penalty0 052108, 2006.

\bibitem{Guta&Jencova}
M. Gu\c{t}\u{a}, A. Jencova
\newblock Local asymptotic normality in quantum statistics
\newblock \emph{Comm. in Math. Phys}, 276:\penalty0 341-379, 2007.



\bibitem{Catana&Bouten&Guta}
C. Catana, M. Gu\c{t}\u{a}, L. Bouten
\newblock  Local asymptotic normality for Quantum Markov Processes
\newblock \emph{ in preparation }


\bibitem{GamLop06}
  \newblock D. Gamerman and H.F. Lopes
  \newblock Markov chain Monte Carlo: stochastic simulation for Bayesian inference
  \newblock CRC Press, 2006




\bibitem{Pritchard}
  \newblock J.K. Pritchard, M.T. Seielstad, A. Perez-Lezaun, and M.W. Feldman. 
  \newblock Population growth of human Y chromosomes: a study of Y chromosome microsatellites. 
  \newblock {\em Mol. Biol. Evol.} 16(12):1791-1798, 1999



\bibitem{Beaumont}
\newblock M.A. Beaumont, W. Zhang, and D.J. Balding.
\newblock Approximate Bayesian Computation in Population Genetics.
\newblock \emph{Genetics}, 162(4):2025-2035,2002.


\bibitem{Marin}
\newblock J.M. Marin, P. Pudlo, C. Robert, and R. Ryder.
\newblock Approximate Bayesian computational methods.
\newblock {\emph Stat. Comput.}, 22(5):1009-1020, 2012.




\bibitem{Schieve}
D.B. Johnson, W.C. Schieve
\newblock Detection statistics in the micromaser
\newblock \emph{Phys. Rev. A}, 63:\penalty0 033808, 2001.

%

%








%


%
%
%
%


 
%



%




  \end{thebibliography}
\end{document}